\documentclass[preprint,12pt]{elsarticle}

\usepackage{lineno}
\usepackage{natbib}
\usepackage{geometry}
\usepackage{fleqn}
\usepackage{graphicx}
\usepackage{newtxtext,newtxmath}
\usepackage[hidelinks]{hyperref}
\usepackage{booktabs}
\usepackage{bm}
\usepackage{amsmath}
\usepackage{enumitem}
\usepackage{esvect}
\usepackage[detect-all]{siunitx} % SI units

\usepackage{framed} % Framing content
\usepackage{color,soul}
\usepackage{subcaption}
\graphicspath{{./figures/}}

\usepackage{multicol}

\journal{Elsevier}
\begin{document}

\begin{frontmatter}

\title{\large{Uncertainty Quantification for Data-Driven Machine Learning Models in Nuclear Engineering Applications: Where We Are and What Do We Need?}}

\author[NCSU]{Xu Wu\corref{mycorrespondingauthor}}
\cortext[mycorrespondingauthor]{Corresponding author}
\ead{xwu27@ncsu.edu}

\author[NCSU,Necsa]{Lesego E. Moloko}
\author[NCSU,Necsa]{Pavel M. Bokov}
\author[NCSU]{Gregory K. Delipei}
\author[NRC]{Joshua Kaizer}
\author[NCSU]{Kostadin N. Ivanov}

\address[NCSU]{Department of Nuclear Engineering, North Carolina State University (NCSU)   \\ 
	Burlington Engineering Laboratories, 2500 Stinson Drive, Raleigh, NC 27695 \\}

\address[Necsa]{The South African Nuclear Energy Corporation SOC Ltd (Necsa)    \\ 
	Building 1900, P.O. Box 582, Pretoria 0001, South Africa \\}

\address[NRC]{Division of Safety Systems, U.S.  Nuclear Regulatory Commission (NRC)    \\ 
	Washington, DC, USA 20555 \\}

\begin{abstract}
	Machine learning (ML) has been leveraged to tackle a diverse range of tasks in almost all branches of nuclear engineering. Many of the successes in ML applications can be attributed to the recent performance breakthroughs in deep learning, the growing availability of computational power, data, and easy-to-use ML libraries. However, these empirical successes have often outpaced our formal understanding of the ML algorithms. An important but under-rated area is uncertainty quantification (UQ) of ML. ML-based models are subject to approximation uncertainty when they are used to make predictions, due to sources including but not limited to, data noise, data coverage, extrapolation, imperfect model architecture and the stochastic training process. The goal of this paper is to clearly explain and illustrate the importance of UQ of ML. We will elucidate the differences in the basic concepts of UQ of physics-based models and data-driven ML models. Various sources of uncertainties in physical modeling and data-driven modeling will be discussed, demonstrated, and compared. We will also present and demonstrate a few techniques to quantify the ML prediction uncertainties. Finally, we will discuss the need for building a verification, validation and UQ framework to establish ML credibility.
\end{abstract}


\end{frontmatter}

\newpage
%%%%%%%%%%%%%%%%%%%%%%%%%%%%%%%%%%%%%%%%%%%%%%%%%%%%%%%%%%%%%%%%%%%%%%%%%%%%%%%%
\section*{Nomenclature}
\label{section:nomenclature}
%%%%%%%%%%%%%%%%%%%%%%%%%%%%%%%%%%%%%%%%%%%%%%%%%%%%%%%%%%%%%%%%%%%%%%%%%%%%%%%%
% \begin{multicols}{2}
	\begin{description}[labelsep=1em, align=left, labelwidth=4em]
	\setlength\itemsep{-0.25em}
	\item[{AI}]     Artificial Intelligence
	\item[{ASME}]   American Society of Mechanical Engineers
	\item[{BE}]     Best Estimate
	\item[{BEPU}]   Best Estimate Plus Uncertainty
	\item[{BNN}]    Bayesian Neural Network
	\item[{CHF}]    Critical Heat Flux
	\item[{CNN}]    Convolutional Neural Network
	\item[{CP}]     Conformal Prediction
	\item[{CSAU}]   Code Scaling Applicability and Uncertainty
	\item[{DE}]     Deep Ensemble
	\item[{DL}]     Deep Learning
	\item[{DNN}]    Deep Neural Network
	\item[{EMDAP}]  Evaluation Models Development and Assessment Procedure
	\item[{GP}]     Gaussian Process
	\item[{i.i.d.}] independent and identically distributed
	\item[{LHS}]    Latin Hypercube Sampling
	\item[{M\&S}]   Modeling and Simulation
	\item[{MCD}]    Monte Carlo Dropout
	\item[{ML}]     Machine Learning
	\item[{MSE}]    Mean Squared Error
	\item[{NEA}]    Nuclear Energy Agency
	\item[{NLL}]    Negative Log Likelihood
	\item[{NNI}]    Neural Network Intelligence
	\item[{NRC}]    U.S. Nuclear Regulatory Commission
	\item[{OECD}]   Organisation for Economic Co-operation and Development
	\item[{QoI}]    Quantity-of-Interest
	\item[{PIML}]   Physics-Informed Machine Learning
	\item[{ReLU}]   Rectified Linear Unit
	\item[{SciML}]  Scientific Machine Learning
	\item[{SRCP}]   Studentized Residual Conformal Prediction
	\item[{SVI}]    Stochastic Variational Inference
	\item[{UQ}]     Uncertainty Quantification
	\item[{VAE}]    Variational AutoEncoder
	\item[{VVUQ}]   Verification, Validation, and Uncertainty Quantification
	\item[{WPRS}]   Working Party on Scientific Issues and Uncertainty Analysis of Reactor Systems
\end{description}
% \end{multicols}

%\linenumbers

%%%%%%%%%%%%%%%%%%%%%%%%%%%%%%%%%%%%%%%%%%%%%%%%%%%%%%%%%%%%%%%%%%%%%%%%%%%%%%%%
\section{Introduction}
\label{section:introduction}
%%%%%%%%%%%%%%%%%%%%%%%%%%%%%%%%%%%%%%%%%%%%%%%%%%%%%%%%%%%%%%%%%%%%%%%%%%%%%%%%

In the past decade, there has been an unprecedented interest in machine learning (ML) among nuclear engineers. ML has been leveraged to tackle a diverse range of tasks in almost all branches of nuclear engineering research. ML is a subset of artificial intelligence (AI) that studies computer algorithms which improve automatically through experience (data). ML algorithms typically build a mathematical model based on training data and then make predictions without being explicitly programmed to do so. Its performance increases with experience; in other words, the machine learns. Deep learning (DL) is a subset of ML that uses deep neural networks (DNNs) to automatically learn representations from data without introducing hand-coded rules or human domain knowledge. DL models' highly flexible architectures can learn directly from large-scale raw data and can increase their predictive accuracy when provided with more data. ML has achieved tremendous success in tasks such as computer vision, natural language processing, speech recognition, and audio synthesis, where the datasets are in the format of \textit{images, text, spoken words and videos}. Meanwhile, their applications in engineering disciplines mostly focus on \textit{scientific data}, which resulted in a burgeoning discipline called \textit{scientific machine learning (SciML)} that blends scientific computing and ML \cite{baker2019basic}. SciML brings together the complementary perspectives of computational science and ML to craft a new generation of ML methods for complex applications. Typical examples for SciML applications include ML-based surrogate modeling for prediction, anomaly detection, inverse problems, uncertainty and sensitivity analysis, physics-informed machine learning (PIML), neural operator for solving ordinary/partial differential equations, etc. SciML is considered an important area of data-driven modeling, which has revolutionized our understanding and discovery of the real world that used to be solely relying on physics-based modeling.

Many of the recent successes in ML applications can be attributed to the modern performance breakthroughs in DL, the growing availability of computational power, data and easy-to-use ML libraries (e.g., scikit-learn, TensorFlow, PyTorch). However, these empirical successes have often outpaced our formal understanding of AI / ML algorithms. One important but underrated area is \textbf{uncertainty quantification (UQ) of ML models}. ML-based models are subject to approximation uncertainty when they are used to make predictions, due to sources including, but not limited to, \textit{data noise, data coverage, extrapolation, imperfect model architecture and the stochastic training process}. Such uncertainty exists even within the training domain because the training data can be sparse and/or noisy. The ML prediction uncertainty is expected to be larger when the amount of data is limited and predictions are made in extrapolated regions.

Compared to the large number of papers available for UQ of physical models, the work on UQ of ML models is relatively limited and new in nuclear engineering. Here by ``physical model'' we mean the mathematical or computational model for a physical system built based on the laws of physics, as a comparison to data-driven models that are built using only data. 
In a recent work \cite{yaseen2023quantification}, we examined UQ methods in DNNs for nuclear engineering applications, focusing on Monte Carlo Dropout (MCD), Deep Ensembles (DE), and Bayesian Neural Networks (BNNs). This paper presented a benchmark comparison of these methods using nuclear simulation datasets, including time-dependent fission gas release and void fraction data. We have noticed discrepancies between the uncertainty bands produced by different methods, and they typically require different DNN architectures and hyperparameters to optimize their performance.
In another recent work \cite{moloko2023prediction}, we quantified the uncertainties introduced by DNNs for axial neutron flux profiles, using MCD and BNN. 
%The neural network models are trained with axial neutron flux measurement data. The ML models are used to make predictions with generalizations (extrapolations) to assemblies and cycles that are not used in the training process. By comparison with the real measurement data, neural network models can predict the axial neutron flux profile very well, while the quantified uncertainties by MCD and BNN are consistent, and envelop the majority of the noisy data points. 
In a follow-up work \cite{moloko2024clustering}, we introduced clustering of the neutron flux profile measurement data in order to improve the regression and UQ results using ML models. 
Xie et al. \cite{xie2024functional} used BNN to quantify the approximation uncertainty when using DNNs as surrogate models in the Bayesian inverse UQ process. 
Abulawi et al. \cite{abulawi2024bayesian} used Bayesian optimized DE for UQ of DNNs, with an application in a system safety case study on sodium fast reactor thermal stratification modeling. 
For UQ of convolutional neural networks (CNNs), Furlong et al. \cite{furlong2024data} used MCD to quantify the CNN approximation uncertainty in predicting crud-induced power shift in a pressurized water reactor.

In the area of turbulence modeling in nuclear thermal-hydraulics, Liu et al. \cite{liu2021uncertainty} performed a comparison of various ML UQ methods for a turbulence model in reactor transient analysis. 
In another work \cite{liu2023uncertainty}, the authors performed UQ of multiphase computational fluid dynamics closure relations using a physics-informed Bayesian approach. 
Very recently, Grogan et al. \cite{grogan2024quantifying} explored DE, MCD and stochastic variational inference (SVI) for quantifying epistemic uncertainties of neural network-based turbulence closures, and potentially how they could be further extended to quantify out-of-training uncertainties. 
Multiple recent studies have been conducted on the UQ of ML for the prediction of critical heat flux (CHF). Mao et al. \cite{mao2024uncertainty} built PIML models for CHF prediction with UQ. 
Kim \cite{kim2024probabilistic} developed a probabilistic neural network approach for CHF prediction. 
In the area of data augmentation using generative ML models, Alsafadi et al. \cite{alsafadi2024predicting} investigated the uncertainty in the synthetic CHF data generated and performed domain generalization analysis using conditional variational autoencoders (VAEs).

In addition to the above-mentioned approaches for UQ of ML, we would like to highlight that a widely used ML technique, the Gaussian Process (GP), has a straightforward measure of its prediction/approximation uncertainty. At each input, the GP ML model will not only provide the mean estimation, but also the variance, also called mean squared error (MSE). Such MSE represents the uncertainty in the GP mean prediction due to various sources including training data coverage and data uncertainty. GP has been widely used in nuclear engineering for many different purposes. For example, GP has been used for surrogate models of expensive physics-based models for forward uncertainty propagation and sensitivity analysis with error predictions and confidence bounds \cite{lockwood2012gradient,iooss2019uncertainty,iooss2019advanced,marrel2022icscream}, as well as inverse UQ \cite{wu2018inverse1,wu2018inverse2,lartaud2023multi}. GP has also been used as an excellent ML model for regression tasks, such as for developing a prognostics approach to nuclear component degradation modeling \cite{baraldi2015prognostics}, surrogate modeling of advanced computer simulations \cite{radaideh2020surrogate}, and for radiation mapping of a nuclear reactor with a mobile robot \cite{west2021use}.

There are also some ongoing efforts in professional organizations that focus on the UQ of ML. The American Society of Mechanical Engineers (ASME) VVUQ 70 subcommittee on ``Verification, Validation, and Uncertainty Quantification (VVUQ) of Machine Learning'' aims to coordinate, promote, and foster the development of standards that provide procedures for assessing and quantifying the credibility of ML algorithms applied to mechanistic and process modeling. The Organisation for Economic Co-operation and Development (OECD) Nuclear Energy Agency (NEA) ``Task Force on AI and ML for Scientific Computing in Nuclear Engineering'', under the Working Party on Scientific Issues and Uncertainty Analysis of Reactor Systems (WPRS), aims at developing benchmark exercises to evaluate the performance of AI/ML in multi-physics Modeling and Simulation (M\&S) of reactor systems \cite{wu2023introducing}. Both of them include a focus on the UQ of SciML models. The OECD/NEA AI/ML Task Force has launched an international benchmark on CHF prediction in October 2023 \cite{lecorre2023benchmark}. The phase 1 exercises include tasks on feature analysis and regression, which have received 47 submissions from 30 institutions \cite{lecorre2025oecd}. Such benchmark analysis on a wide range of different ML models will support the development and performance assessment of ML methods and build communities of practice to exchange know-how on AI and ML applications. The phase 2 exercises will include a major task on UQ of ML.

The goal of this paper is to clearly explain and illustrate the importance of UQ of ML models, or more specifically, SciML models, since ML has been mainly used for scientific computing in nuclear engineering. We will elucidate the differences in the basic concepts of UQ of physics-based models and data-driven ML models. Various sources of uncertainty in physical modeling and ML will be discussed and explained. We will present some of the available techniques to quantify the prediction/approximation uncertainty in ML models. We will also demonstrate the UQ of SciML models through a simple test example and a realistic nuclear engineering example. Finally, we will discuss the need for building a verification, validation and UQ (VVUQ) framework to establish ML credibility.

The structure of this paper is as follows. Section \ref{section:uncertainties} provides a detailed definition of the uncertainties of physical modeling as well as data-driven modeling. In Section \ref{section:methodologies}, we provide a brief overview of the existing approaches that can be used to quantify the prediction uncertainties in ML models. Section \ref{section:demonstration} presents the demonstration examples on UQ of ML. A discussion of the VVUQ needs of ML models is included in Section \ref{section:discussions}. Finally, Section \ref{section:conclusions} concludes this paper.

%%%%%%%%%%%%%%%%%%%%%%%%%%%%%%%%%%%%%%%%%%%%%%%%%%%%%%%%%%%%%%%%%%%%%%%%%%%%%%%%
\section{Definition of Uncertainties}
\label{section:uncertainties}
%%%%%%%%%%%%%%%%%%%%%%%%%%%%%%%%%%%%%%%%%%%%%%%%%%%%%%%%%%%%%%%%%%%%%%%%%%%%%%%%

%%%%%%%%%%%%%%%%%%%%%%%%%%%%%%%%%%%%%%%%%%%%
\subsection{Types of uncertainties}
\label{section:uncertainties-type}
%%%%%%%%%%%%%%%%%%%%%%%%%%%%%%%%%%%%%%%%%%%%

Uncertainties in physics-based M\&S can generally be categorized as three types: \textit{aleatory, epistemic and mixed}. Even though many people in the nuclear area are very familiar with these concepts, we feel it is necessary to reiterate here, because later they will potentially create confusion with the definitions used in the ML area. 

Aleatory uncertainty arises from randomness, such as stochastic variations in the physical system. It is irreducible, in that better data or improved models cannot reduce it. A property of aleatory uncertainty is that the randomness can be described by a distribution. For example, it can be represented by continuous/discrete random variables such as normal, uniform, etc. 

Epistemic uncertainty arises from lack of knowledge, particularly of the physical model parameters and imperfections in the mathematical models. It is reducible by more data or information. One does not describe it with a probability distribution. In many cases, the best one can do is to bound the uncertainty -- we deal with intervals instead of probabilities. Therefore, epistemic uncertainty can be represented by continuous/discrete intervals, discrete sets, etc. 

Note that in the Bayesian UQ formulation, we usually do not make a distinction between epistemic and aleatory uncertainties \cite{oden2010computer}. In this case, probability represents our confidence in some proposition, given all currently available information. Bayesian inference gives us the formalism to update that confidence (probability) when new information becomes available. Also, it is worth noting that it is not uncommon in BEPU (Best Estimate Plus Uncertainty) studies that researchers use a probabilistic representation of all the uncertainties (aleatory and epistemic). See a debate in \cite{der2009aleatory}.

%%%%%%%%%%%%%%%%%%%%%%%%%%%%%%%%%%%%%%%%%%%%
\subsection{Sources of uncertainties in physics-based M\&S}
\label{section:uncertainties-physical}
%%%%%%%%%%%%%%%%%%%%%%%%%%%%%%%%%%%%%%%%%%%%

The nuclear community has been studying the UQ of physical models since the late 1980s, when the best-estimate (BE) safety analysis strategy started to be embedded in the Code Scaling Applicability and Uncertainty (CSAU) \cite{boyack1990quantifying}, and Evaluation Models Development and Assessment Procedure (EMDAP) \cite{kaizer2018the}. This strategy is commonly referred to as Best Estimate plus Uncertainty (BEPU) \cite{d2012best,wilson2013historical} that has been accepted by the U.S. Nuclear Regulatory Commission (NRC). The concept of UQ in the nuclear community generally means forward UQ, which is the process of propagating uncertainties from input parameters to the output quantities-of-interest (QoIs). Numerous methodologies have been developed for forward UQ of different types of problems.

Forward UQ requires knowledge of the input uncertainties, which has been usually specified using ``expert opinion'' or ``user self-evaluation''. Another area of UQ that has become popular in the last decade is inverse UQ \cite{wu2021comprehensive}, which is the process to inversely quantify the input uncertainties while keeping model outputs consistent with measurement data.

\begin{figure}[!ht]
	\centering
	\includegraphics[width=.99\linewidth]{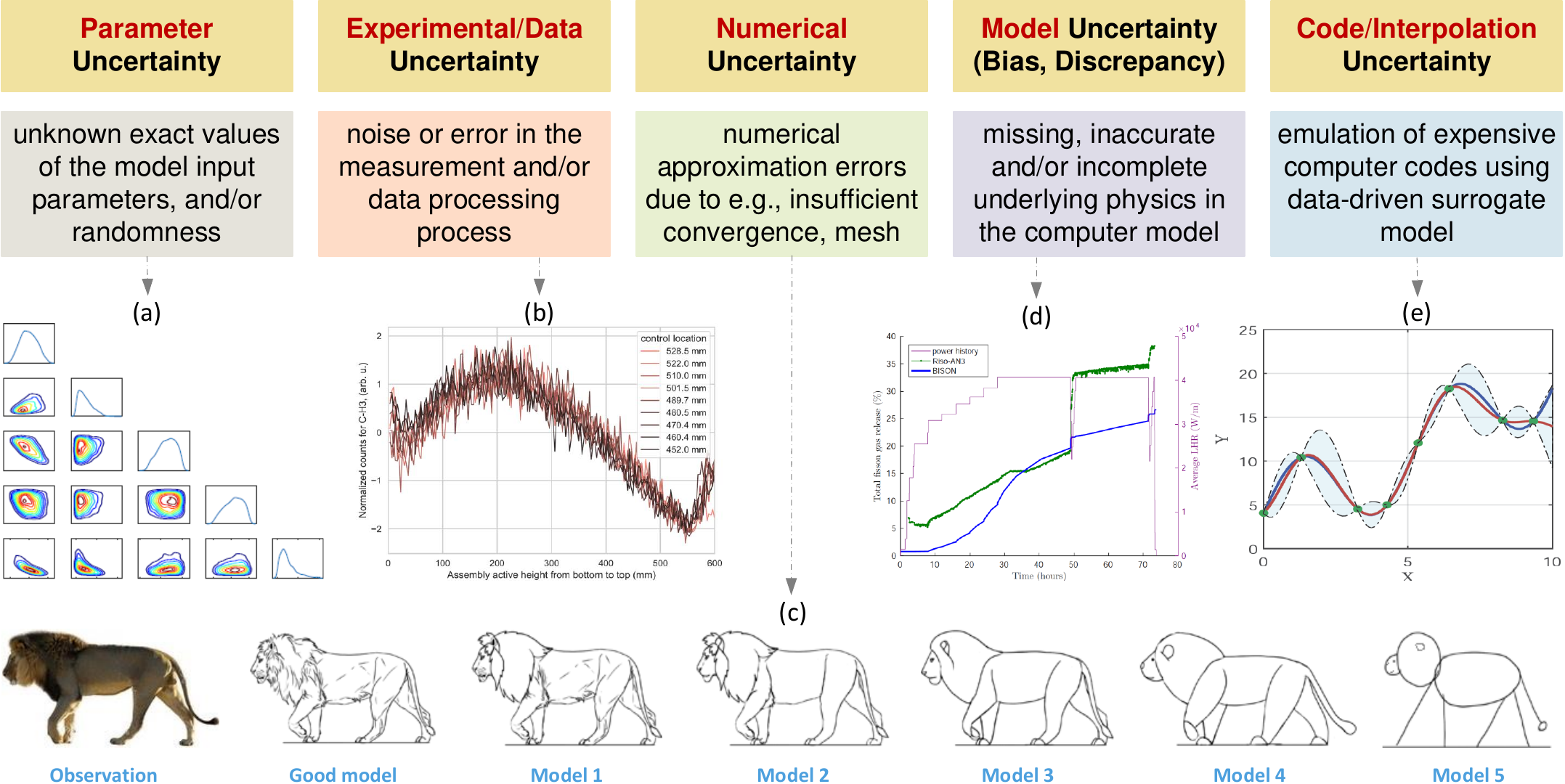}
	\caption{Illustration of uncertainty sources in physics-based M\&S.}
	\label{figure:uncertainty-source-physical-model}
\end{figure}

Figure \ref{figure:uncertainty-source-physical-model} illustrates the various sources of uncertainties in physics-based M\&S. These sources of uncertainties include most, if not all, of the quantifiable uncertainties that should be considered in M\&S. Failing to account for any of these uncertainties will result in biased predictions and uncertainty estimations.

%-------------------------------------------
\subsubsection{Parameter uncertainty}
%-------------------------------------------

The parameter uncertainty is due to ignorance in the exact values of input parameters (epistemic) or randomness (aleatory). Figure \ref{figure:uncertainty-source-physical-model}(a) shows the marginal distribution of five input parameters as well as their pairwise correlations. Examples of such parameters in nuclear reactors M{\&}S are the nuclear data, physical model parameters, and material properties. This is the most common source of uncertainties considered in UQ.

%-------------------------------------------
\subsubsection{Experiment/data uncertainty}
%-------------------------------------------

The experimental uncertainty is due to noise or error in the measurement process, and/or data processing. Figure \ref{figure:uncertainty-source-physical-model}(b) shows the noise data in the SAFARI-1 axial neutron flux profiles \cite{moloko2023prediction}. It is generally expected to be reported along with the measurement data. If not, it can be evaluated in certain cases, for example, through inverse UQ using full Bayesian inference.

%-------------------------------------------
\subsubsection{Numerical uncertainty}
%-------------------------------------------

The numerical uncertainty is due to numerical approximation errors from, e.g., convergence, truncation, mesh, etc. Figure \ref{figure:uncertainty-source-physical-model}(c) uses a cartoon to show a few ``models'' with different levels of refinement, which explains how insufficient mesh can lead to large numerical uncertainty. Approaches such as Richardson extrapolation can be used to estimate this uncertainty \cite{oberkampf2010verification}. Note that in the literature, it is quite common to treat this source as a part of the model uncertainty below, since they are usually not separable from each other.

%-------------------------------------------
\subsubsection{Model uncertainty}
%-------------------------------------------

The model uncertainty is due to missing, inaccurate and/or incomplete underlying physics phenomena included in the computer models. There is no unanimous definition of model uncertainty in the literature; sometimes numerical uncertainty is treated as part of the model uncertainty. Model uncertainty is also called model bias/error/discrepancy/inadequacy, or model form uncertainty in the literature. Figure \ref{figure:uncertainty-source-physical-model}(d) shows that incomplete physics in a fuel performance code that cannot fully capture the burst release of fission gas during a power transient.

%-------------------------------------------
\subsubsection{Code/interpolation uncertainty}
\label{section:uncertainties-code}
%-------------------------------------------

The code/interpolation uncertainty is due to the emulation of computationally prohibitive codes with surrogate models (also called metamodels, response surfaces, or emulators). Surrogate models are approximations of the input/output relation of a physics-based computer model. They are built from a limited number of full model runs (training set) and a learning algorithm, and they can be evaluated very fast. In other words, the ML-based surrogate model essentially ``interpolates'' the full model using the training dataset. Figure~\ref{figure:uncertainty-source-physical-model}(e) uses a GP ML model to show that the surrogate model can have uncertainty when the training data does not have a good coverage of the training domain. Note that the code/interpolation uncertainty does not exist if the full model is used. Here, the full model means the original physical model/code, as a comparison to the surrogate model.

%%%%%%%%%%%%%%%%%%%%%%%%%%%%%%%%%%%%%%%%%%%%
\subsection{Sources of uncertainties in data-driven ML models}
\label{section:uncertainties-ML}
%%%%%%%%%%%%%%%%%%%%%%%%%%%%%%%%%%%%%%%%%%%%

\begin{figure}[!ht]
	\centering
	\includegraphics[width=.99\linewidth]{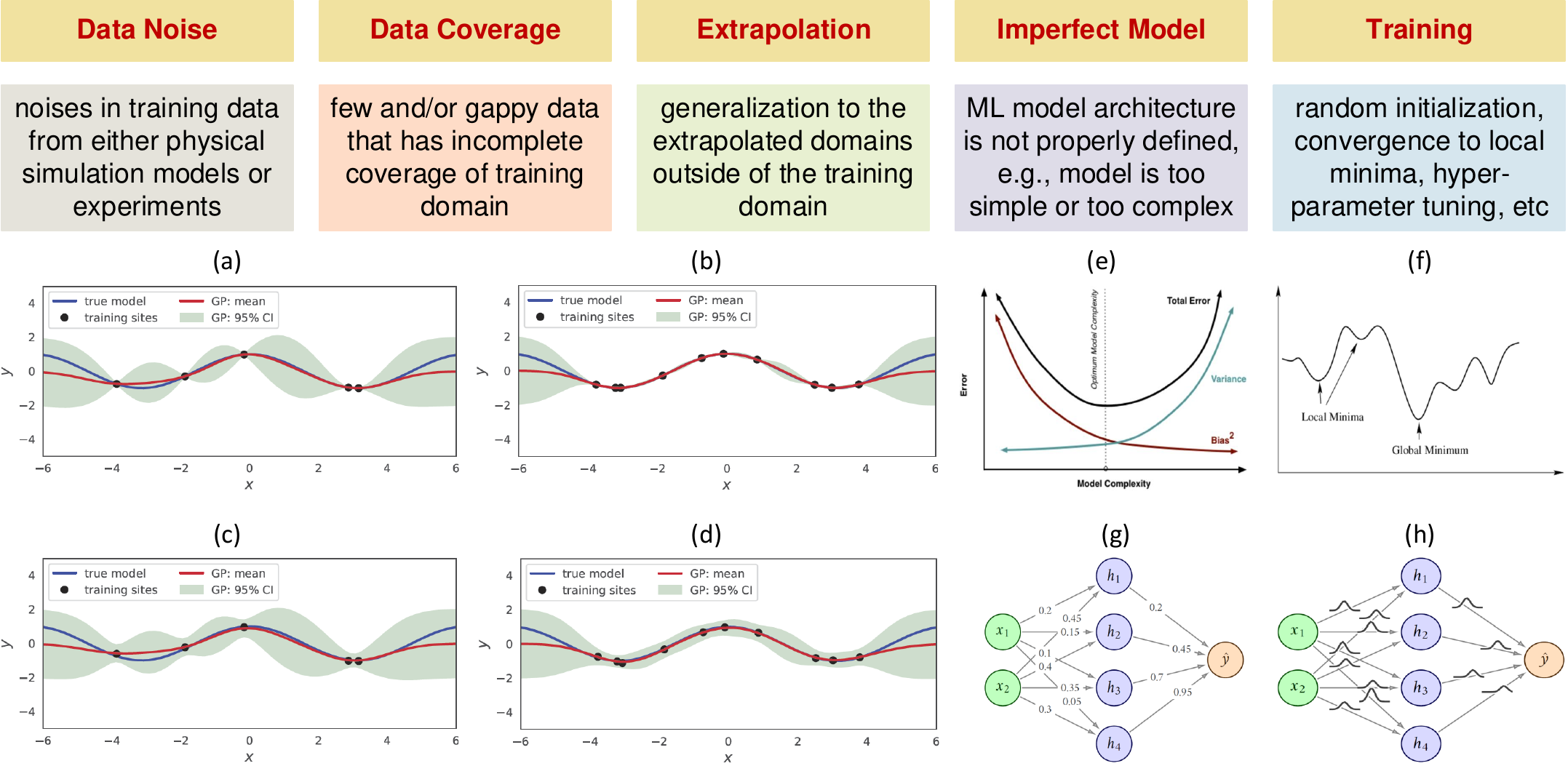}
	\caption{Illustration of uncertainty sources in data-driven ML models.}
	\label{figure:uncertainty-source-data-driven-ML}
\end{figure}

ML-based models are subject to approximation uncertainties when they are used to make predictions. Such uncertainty exists even within the training domain because the training data can be sparse and/or noisy. We summarize the sources of uncertainties for the ML models in five categories, as illustrated in Figure \ref{figure:uncertainty-source-data-driven-ML}. Unlike the uncertainties in physics-based modeling, these sources are usually not well-separated and independent from each other. Therefore, it is usually only possible to quantify them together. The total uncertainty (combined effects of the five sources in Figure \ref{figure:uncertainty-source-data-driven-ML}) in ML approximation/prediction is the ``\textbf{code/interpolation uncertainty}'' in Figure \ref{figure:uncertainty-source-physical-model}. 

Note that in Section \ref{section:uncertainties-code}, we mentioned that the code/interpolation uncertainty is a consequence of building surrogate models using physical model simulation data. The discussion in this section also applies to the case when an ML model is trained using experimental data. Instead of defining a new terminology, we will keep using code/interpolation uncertainty since, in this case, the ML model is essentially ``interpolating'' the experimental data. Figures \ref{figure:uncertainty-source-data-driven-ML}(a) - (d) use GP-based ML models that are trained on 5 points in (a) and (c), and 10 points in (b) and (d), respectively. The training data points in (a) and (b) are deterministic, while the training data points in (c) and (d) have noise.

\vspace{1em}

%-------------------------------------------
\subsubsection{Data noise}
%-------------------------------------------

Noise in training data from either physical simulation models or experiments can lead to ML uncertainty. By comparing Figures \ref{figure:uncertainty-source-data-driven-ML}(a) and \ref{figure:uncertainty-source-data-driven-ML}(c), and Figures \ref{figure:uncertainty-source-data-driven-ML}(b) and \ref{figure:uncertainty-source-data-driven-ML}(d), when the training data has uncertainty, the ML prediction is also uncertain, as shown in the light green shaded area. This happens even when the data points cover the training domain sufficiently in Figure \ref{figure:uncertainty-source-data-driven-ML}(d).

%-------------------------------------------
\subsubsection{Data coverage}
%-------------------------------------------

Few and/or gappy data that has incomplete coverage of training domain can cause ML uncertainty. By comparing Figures \ref{figure:uncertainty-source-data-driven-ML}(a) and \ref{figure:uncertainty-source-data-driven-ML}(b), and Figures \ref{figure:uncertainty-source-data-driven-ML}(c) and \ref{figure:uncertainty-source-data-driven-ML}(d), when the training data size is small, an insufficient coverage of the training domain will cause large ML prediction uncertainty.

%-------------------------------------------
\subsubsection{Extrapolation}
%-------------------------------------------

Generalization of ML models to extrapolated domains outside the training domain (also known as domain generalization) can result in large uncertainties. This is one of the major concerns in SciML applications. By looking at any one of Figures \ref{figure:uncertainty-source-data-driven-ML}(a) - \ref{figure:uncertainty-source-data-driven-ML}(d), when GP models are extrapolated outside of the training domain (beyond the left and right bounds), the error between the mean function and true function increases and prediction uncertainty quickly increases as well.

%-------------------------------------------
\subsubsection{Imperfect model}
%-------------------------------------------

When an ML model's architecture is not properly defined, e.g., the model is too simple or too complex with respect to the data, the ML prediction can have a large error/uncertainty. As shown in Figure \ref{figure:uncertainty-source-data-driven-ML}(e), the total error in ML models does not always decrease when the model complexity increases. Generally, a good balance between bias and variance should be reached, which is referred to as the ``bias-variance tradeoff''. For example, when the training data is limited, a DNN models with many hidden layers and hidden neurons in each layer, and many input features will not be well-trained, leading to large prediction uncertainty.

%-------------------------------------------
\subsubsection{Stochastic training process}
%-------------------------------------------

Issues such as random initialization, convergence to local minima, hyper-parameter tuning, posterior inference, etc. can result in ML uncertainty. Figure \ref{figure:uncertainty-source-data-driven-ML}(f) shows that the loss function might be multi-modal, with a global minimum and many local minima. Random initialization of the ML model parameters (e.g. weights and biases in a DNN), and stochastic gradient descent training process can cause the training to converge to different local minima. Furthermore, in the case of a BNN (Figure \ref{figure:uncertainty-source-data-driven-ML}(h)), where the parameters follow distributions, instead of a regular neural network (Figure \ref{figure:uncertainty-source-data-driven-ML}(g)), where the parameters are deterministic, the BNN parameters are estimated by Bayesian inference, which can result in different posterior distributions of neural network parameters. These factors can all lead to ML prediction uncertainty as well.

%%%%%%%%%%%%%%%%%%%%%%%%%%%%%%%%%%%%%%%%%%%%
\subsection{Connections between UQ and SciML}
\label{section:uncertainties-UQ-and-SciML}
%%%%%%%%%%%%%%%%%%%%%%%%%%%%%%%%%%%%%%%%%%%%

\begin{figure}[!ht]
	\centering
	\includegraphics[width=.99\linewidth]{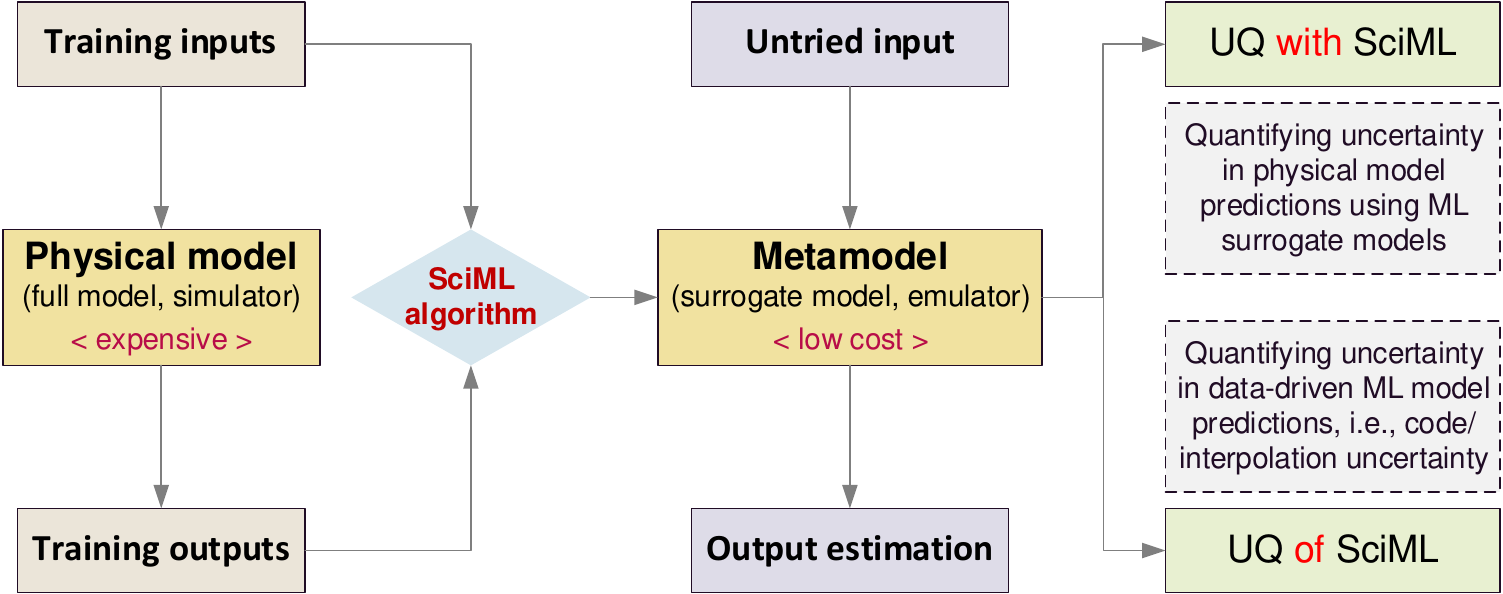}
	\caption{Connections between UQ and SciML.}
	\label{figure:connection-UQ-and-SciML}
\end{figure}

Based on the above discussion, we summarize the connections between UQ and SciML as two-fold: (1) UQ with SciML, and (2) UQ of SciML, as illustrated in Figure \ref{figure:connection-UQ-and-SciML}.

\textbf{UQ with SciML} means using ML-based surrogate models to quantify the uncertainties in physical model predictions. This is motivated by the fact that many approaches for uncertainty, sensitivity, calibration and optimization studies require a large number of physical model simulations, usually at different inputs. For example, Monte Carlo sampling and its variations (simple/crude Monte Carlo, Latin hypercube sampling -- LHS, maximin LHS, and low discrepancy sequences), stochastic spectral methods (polynomial chaos expansion, stochastic collocation), etc. Additionally, the uncertainty propagation of both epistemic and aleatoric sources requires nested approaches that further increase the number of code evaluations \cite{roy2011comprehensive}. Building an accurate and fast-running surrogate model using ML techniques based on only a limited number of full model runs can significantly reduce the computational time. 

As explained in Section \ref{section:uncertainties-code}, the surrogate modeling process using ML techniques can introduce code/interpolation uncertainty, which is the motivation of \textbf{UQ of SciML} models. This is the step to quantify the extra uncertainty introduced by using the ML-based surrogate models, instead of the full physics-based model itself, by considering all the uncertainty sources in Section~\ref{section:uncertainties-ML} together. Inclusion of code/interpolation uncertainty is important for a comprehensive UQ of physical models, because this uncertainty may be large due to the reasons explained in Section~\ref{section:uncertainties-ML}. However, we noticed that in the majority of previous works on UQ of physics-based models in nuclear engineering that used ML-based surrogate models, the code/interpolation uncertainty was not considered, especially when DNNs are used. In other words, very few works (e.g. \cite{Labarile2018, Bokov2018Table, Bokov2022:Hierarchical}) have considered ``UQ of SciML'' when performing ``UQ with SciML''. One exception is when GP is used, because the prediction uncertainty in a GP model is inherently available. A literature survey on the ML work with UQ considered has been provided in the introduction section.

%For the case of UQ with SciML, the uncertainty sources mentioned previously can be represented by the following equation:
%\begin{equation} \label{eq:uq}
%	Y_\mathrm{real}=Y_\mathrm{obs}+\epsilon=F(X;p)+\delta_F(X;p) = f_{\theta}(X;p\mid\theta) + \delta_f(X;p\mid\theta),   
%\end{equation}
%where $Y_\mathrm{real}$ is the true unknown value of a measured quantity, $Y_\mathrm{obs}$ is the measured value of the quantity, $\epsilon$ is the experimental uncertainty, $\delta_F$ is the uncertainty due to models and numerical approximations, $X$ are the uncertain measurement conditions, and $p$ are the uncertain parameters for the numerical model $F$. When a ML model $f_{\theta}$ is used as a surrogate to $F$, then additional uncertainties are added due to the estimation of the model parameters $\theta$ (imperfect model, stochastic training) and due to the model form uncertainty $\delta_f$ that incorporates uncertainties from the data noise, data coverage and extrapolation. \TODO{Gregory: add more details}

%%%%%%%%%%%%%%%%%%%%%%%%%%%%%%%%%%%%%%%%%%%%%%%%%%%%%%%%%%%%%%%%%%%%%%%%%%%%%%%%
\section{Methodologies for UQ of ML}
\label{section:methodologies}
%%%%%%%%%%%%%%%%%%%%%%%%%%%%%%%%%%%%%%%%%%%%%%%%%%%%%%%%%%%%%%%%%%%%%%%%%%%%%%%%

As DNNs have the greatest potential for SciML applications in nuclear engineering, in this section we will provide a brief overview of the existing techniques (MCD, DE and BNN) for UQ of DNNs. More details on the theory and demonstration examples can be found in our previous work \cite{yaseen2023quantification,moloko2023prediction}. Note that more details are presented on Conformal Prediction (CP) because it can be applied for UQ of any type of ML model.

%%%%%%%%%%%%%%%%%%%%%%%%%%%%%%%%%%%%%%%%%%%%
\subsection{Monte Carlo Dropout}
\label{sec:methodologies-MCD}
%%%%%%%%%%%%%%%%%%%%%%%%%%%%%%%%%%%%%%%%%%%%

\begin{figure}[!ht]
	\centering
	\includegraphics[width=.8\linewidth]{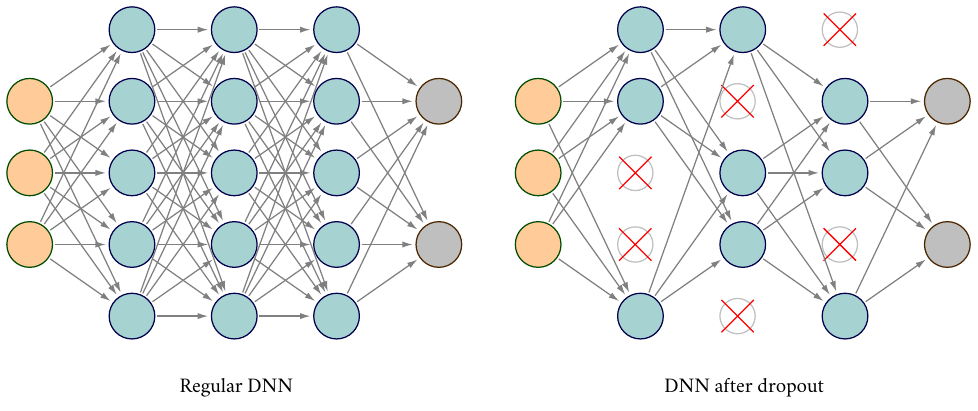}
	\caption{Illustration of MCD.}
	\label{figure:method-MCD}
\end{figure}

MCD is one of the most widely used methods to quantify DNN prediction uncertainties \cite{gal2016dropout}. Dropout is a regularization approach to avoid over-fitting by randomly ignoring or dropping out certain hidden layer neurons during the training of DNNs \cite{srivastava2014dropout}, as illustrated in Figure~\ref{figure:method-MCD}. Dropout for regularization is used during DNN training but not in prediction. With MCD, dropout is also used when making predictions in which the hidden neurons in DNNs are randomly turned off in the prediction step. In this way, the network can be evaluated for the same input multiple times, resulting in a collection of predictions that can be used to estimate the uncertainties in the predictions.

\vspace{0.5em}

%%%%%%%%%%%%%%%%%%%%%%%%%%%%%%%%%%%%%%%%%%%%
\subsection{Deep Ensemble}
\label{sec:methodologies-DE}
%%%%%%%%%%%%%%%%%%%%%%%%%%%%%%%%%%%%%%%%%%%%

\begin{figure}[!ht]
	\centering
	\includegraphics[width=.8\linewidth]{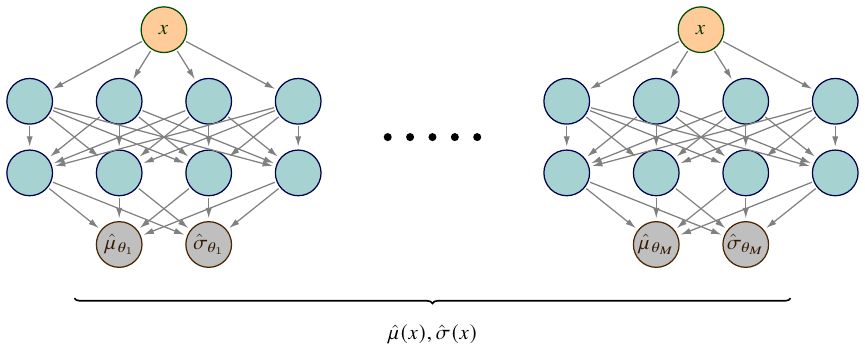}
	\caption{Illustration of DE.}
	\label{figure:method-DE}
\end{figure}

Another method is DE \cite{lakshminarayanan2017simple}, which assumes that the training data has a given parameterized distribution where the distribution parameters depend on the inputs. For example, if one assumes that the training data outputs follow a Gaussian distribution, the distribution parameters, mean value and variance, can fully characterize the Gaussian distribution. The neural network will no longer predict the outputs in the training set, but instead will predict the distribution parameters (the mean value and standard deviation in the case of Gaussian), as a function of the training inputs. The distribution parameters are directly learned by the DNN outputs. For additional robustness, we will not only train one network but an ensemble of networks with the same architecture but different random initializations and training processes, as shown in Figure~\ref{figure:method-DE}. For predictions, the distributions from all the individual networks are then averaged for a final estimate, resulting in a mixture distribution that can be analyzed for the prediction uncertainty.

\vspace{0.75em}

%%%%%%%%%%%%%%%%%%%%%%%%%%%%%%%%%%%%%%%%%%%%
\subsection{Bayesian Neural Network}
\label{sec:methodologies-BNN}
%%%%%%%%%%%%%%%%%%%%%%%%%%%%%%%%%%%%%%%%%%%%

\begin{figure}[!ht]
	\centering
	\includegraphics[width=.8\linewidth]{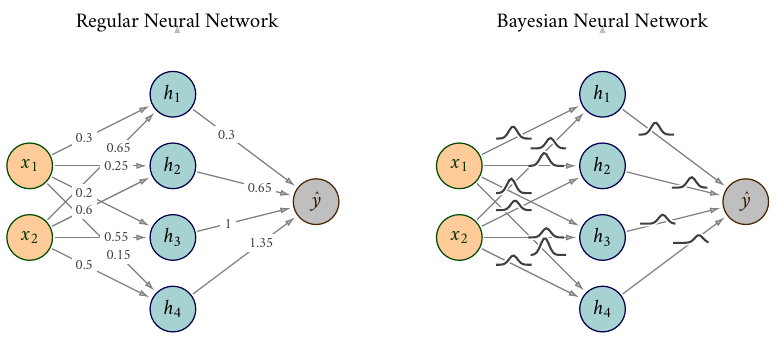}
	\caption{Illustration of BNN.}
	\label{figure:method-BNN}
\end{figure}

UQ of DNN models can also be achieved through the BNN \cite{blundell2015weight,goan2020bayesian}, which is a neural network with distributions over parameters (weights and bias). In a regular neural network, we obtain deterministic estimations of the network parameters. Consequently, when an input is presented to the neural network, we get a deterministic output. In BNNs, a prior distribution is specified upon the parameters of the neural networks and then, given the training data, the posterior distributions over the parameters are computed, as shown in Figure \ref{figure:method-BNN}. As a result, when an input is fed to the trained BNN, the prediction will also be a distribution instead of a deterministic value. The prediction uncertainty is therefore directly available from BNN. Since exact Bayesian inference is computationally intractable for neural networks, variational inference \cite{blei2017variational} has been used to find the posterior distributions of the parameters.

%%%%%%%%%%%%%%%%%%%%%%%%%%%%%%%%%%%%%%%%%%%%
\subsection{Gaussian Process}
\label{sec:methodologies-GP}
%%%%%%%%%%%%%%%%%%%%%%%%%%%%%%%%%%%%%%%%%%%%

Depending on the context, the term GP refers to either (a) stochastic processes such that every finite collection of those random variables has a multivariate normal distribution or (b) a statistical modeling and nonparametric supervised ML method, often used for regression, classification, and optimization tasks. In the latter case, GP modeling is also called Kriging or spatial correlation modeling. It was originally developed by geologists in the 1950s to predict distribution of minerals over an area of interest given a set of sampled sites. 
It was made popular in the context of modeling and optimization by Sacks et al. \cite{sacks1989design} and Jones et al. \cite{jones1998efficient} respectively. GP has been widely used to construct surrogate models for computer models in many areas, see detailed reviews in \cite{kleijnen2009kriging} and \cite{martin2005use}.

A GP model can be considered as a generalized linear regression model that accounts for the correlation in the residuals between the regression model and the observations. Here ``observations'' stand for the samples in the training dataset, which can be from either computational results or experimental data. The difference between GP and a linear regression model is that GP assumes that the model response is continuous and smooth over the input domain, which is true for most computer models. We thus believe that if two points are close to each other in the input domain, the residuals in the regression model should be close. It follows that GP does not treat the residuals as independent, assuming that the correlation between the residuals are related to the distance between the corresponding input points. Such correlation can be represented using spatial correlation kernels \citep{Rasmussen2005Gaussian}. By definition of a GP, the joint distribution of the function's value at a finite number of input points is a multivariate normal distribution. This allows us to fit models from a finite number of observed data points and make predictions for finitely many new data points by applying rules for conditioning Gaussians.

\begin{figure}[!htb]
	\centering
	\includegraphics[width=.95\linewidth]{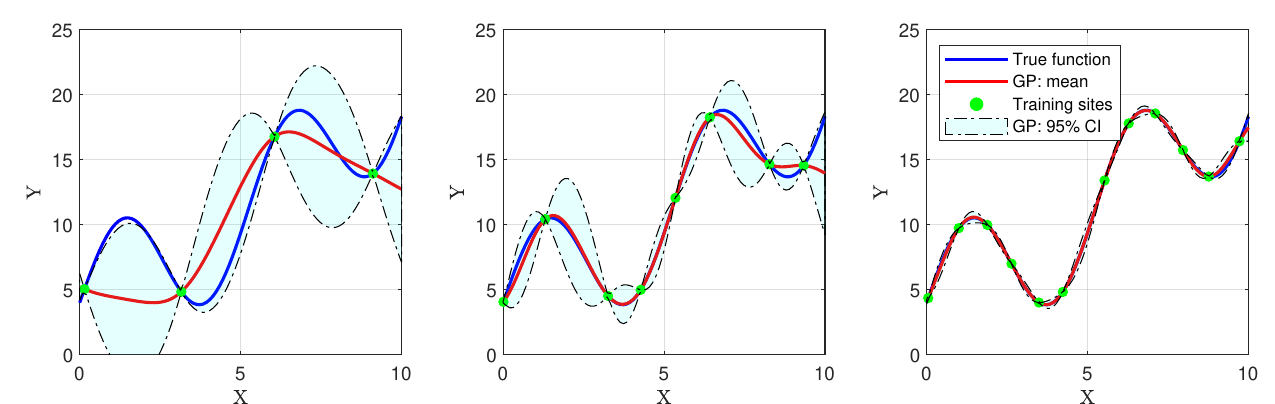}
	\caption{Illustration of GP prediction mean and uncertainty (95\% confidence interval is presented, i.e., twice the standard deviation $\sigma$ at every $x$).}
	\label{figure:method-GP}
\end{figure}

The greatest advantage of GP is that the prediction/approximation uncertainty is inherently available. The GP ML model offers not only the mean prediction at any input but also the variance or MSE associated with the mean prediction. More details on the mathematical definitions of GP can be found in our previous work \cite{wu2018inverse1}. Figure \ref{figure:method-GP} illustrates the way GP approximates a test function based on a limited number of training samples and the associated uncertainties. At every input, the GP estimation follows a Gaussian distribution with a mean prediction and a variance. The variance of the prediction decreases as the test point gets closer to training points. When the number of training points increases, the accuracy of the approximation improves quickly. Note that, as shown in Figure \ref{figure:uncertainty-source-data-driven-ML}, when the training data is uncertain, GP can be trained in a way not interpolating the training samples but has non-zero uncertainty everywhere.

%%%%%%%%%%%%%%%%%%%%%%%%%%%%%%%%%%%%%%%%%%%%
\subsection{Conformal Prediction}
\label{sec:methodologies-CP}
%%%%%%%%%%%%%%%%%%%%%%%%%%%%%%%%%%%%%%%%%%%%

In recent years, the CP method has attracted interest for quantifying the prediction uncertainty of ML models in the ML community \cite{shafer2008tutorial, angelopoulos2023conformal}. To the authors' knowledge, there have been no applications of CP for UQ (either physics-based or data-driven) in nuclear engineering problems. Therefore, in this section we will provide a relatively more detailed and self-contained introduction on CP. CP aims at assigning confidence intervals in different model prediction regions or sets using rank statistics. It can be used to quantify uncertainty in the predictions made by any model or algorithm, from either physical models or data-driven ML models. Therefore, we expect the CP method to become widely accepted by ML practitioners in the nuclear engineering area for UQ of ML models.

For the case of a response of interest $Y \in \mathcal{Y}$, let us assume that there is an available set of independent and identically distributed (i.i.d.) values $\mathcal{D}_n=\left\{Y_i\right\}^n_{i=1}$. Rank statistics use the fact that any additional i.i.d. response $Y_{n+1}$ will be located with a uniform probability in any of the subintervals created by ranking the values in $D_n$ in increasing order $\mathcal{D}^r_n=\left\{Y^r_i\right\}^n_{i=1}$, where $Y^r_{i}\leq Y^r_{i+1}$. The first subinterval is $(-\infty, Y^r_1]$, the last subinterval is $[Y^r_{n}, \infty)$, and the in-between subintervals are $\left[Y^r_{i}, Y^r_{i+1}\right]^{n-1}_{i=1}$. This creates a total of $n+1$ subintervals and thus $Y_{n+1}$ will have a probability of $\frac{1}{n+1}$ to be within any of them. Using this property, it is possible to find a prediction interval $\hat{C}_{n,\alpha}$ for the response $Y_{n+1}$ that will satisfy a selected confidence level $\alpha$ through:
\begin{align}\label{eq:cp_definition}
	\mathbb{P} \left( Y_{n+1} \in \hat{C}_{n,\alpha} \right) \geq 1 - \alpha.
\end{align}

In the case of a one-sided interval, the prediction interval is $\hat{C}_{n,\alpha} = (-\infty, \hat{q}_{n,\alpha}]$. The $\hat{q}_{n,\alpha}$ is the adjusted empirical quantile obtained through:
\begin{align}\label{eq:cp_quantile}
	\hat{q}_{n,\alpha} = \hat{q}_{n} \left( \frac{\lceil (1-\alpha)(n+1) \rceil }{n} \right),
\end{align}
where $\hat{q}_{n}$ is the empirical quantile of $Y$ constructed using the $\mathcal{D}_n$ data, and $\lceil x \rceil$ is the ceil function that provides the closest integer $u$ for which $u \geq x$. If we assume that the integer found is $u$, then equation~\eqref{eq:cp_quantile} essentially defines the desired quantile as $\hat{q}_{n,\alpha}=Y^r_u$ and thus:
\begin{align}\label{eq:cp_interval}
	\mathbb{P} \left( Y_{n+1} \leq Y^r_u \right) \geq 1 - \alpha .
\end{align}

% To better understand this we provide a simple example where we assume a total of $n=100$ observations $(Y_I)^{100}_{i=1}$ are available. We select a confidence level $\alpha=0.05$ and desire to find the appropriate interval. The first step is ranking the obtained values into $(Y^r_i)^{100}_{i=1}$. The second step is computing the adjusted empirical quantile  $\hat{q}_{n,\alpha}$. To do this we first compute the integer $u=\lceil (1-0.05)(101) \rceil = 96$. This means that $\hat{q}_{n,\alpha} = \hat{q}_{n}(0.96) = Y^r_{96}$. The third step is to set the corresponding prediction interval as: 
% %
% \begin{align}
% 	\mathbb{P} \left( Y_{101} \leq Y^r_{96} \right) \geq 0.95
% \end{align}

% We can even compute manually the exact probability since we now that $Y_{101}$ will have a $\frac{1}{101}$ chance to be in each subinterval. This means that the following probabilities hold true:
% \begin{align*}\label{eq:cp_interval_example_res}
% 	\mathbb{P} \left( Y_{101} \leq Y^r_{1} \right) &= \frac{1}{101}, \quad \mathbb{P} \left( Y_{101} \leq Y^r_{2} \right) = \frac{2}{101}, \\
% 	&\, \, \, \vdots \\
% 	\mathbb{P} \left( Y_{101} \leq Y^r_{95} \right) &= \frac{95}{101}=0.941, \quad \color{red}{\mathbb{P} \left( Y_{101} \leq Y^r_{96} \right) = \frac{96}{101}=0.9505} \color{black}, \\ 
% 	&\, \, \, \vdots \\
% 	\mathbb{P} \left( Y_{101} \leq Y^r_{100} \right) &= \frac{100}{101}=0.99 \quad \mathbb{P} \left( Y_{101} < \infty \right) = \frac{101}{101}=1.0
% \end{align*}

This rank statistics property holds for any distribution $\mathbb{P}$ and for any finite sample size $n$, rendering it very general. The main requirement is that $\left\{Y_i\right\}^{n+1}_{i=1}$ are exchangeable, meaning that the ordering of the values in this set does not impact the distribution $\mathbb{P}$, something that is even less strict than i.i.d.  

% We define an input space of independent variables or features $X \in \mathcal{X}$, and the corresponding output space $Y \in \mathcal{Y}$. We assume that a set of $n$ observations are available $D_n=(X_i, Y_i)^n_{i=1}$.In the context of this work, we can further assume that $\mathcal{X}=\mathbb{R}^d$, with $d$ the number of input features, and that $\mathcal{Y}=\mathbb{R}$ resulting in the prediction of a scalar output.
CP expands this general property to ML model predictions for the dependent variable $Y$. We consider that a ML model $\hat{f}: \mathbb{R}^d \rightarrow \mathbb{R}$ is available that can provide point predictions $\hat{Y_i}=\hat{f}(X_i)$. Similarly to rank statistics, the goal of CP is to provide a prediction interval $\hat{C}_{n,\alpha}(X)$ with a selected confidence level $\alpha$ for a new data point $\left \{X_{n+1}, \hat{Y}_{n+1}\right \}$. It is important to highlight that the true $Y_{n+1}$ is unknown and only the model prediction $\hat{Y}_{n+1}$ is available. The goal is thus to provide the interval $\hat{C}_{n,\alpha}(X)$ that will cover the true value $Y_{n+1}$ with a probability of at least $(1-\alpha)$. To achieve this, CP uses a nonconformity score $L_i=L(X_i,Y_i;\hat{f}), \; i=1,2,\dots,n$. This score measures how expected is the pair $\left \{X_i,Y_i \right \}$ based on the prediction of $\hat{f}(X_i)$, with small values indicating strongly expected predictions (i.e. in agreement with the model assessment) while large values indicate strongly unexpected predictions (i.e. in disagreement with the model assessment). The resulting prediction interval is defined as:
\begin{align}
	\hat{C}_{n, \alpha}(X_{n+1}) = \left \{ y: L(X_{n+1},y) \leq \hat{q}_n \left( \frac{\lceil (1-\alpha)(n+1) \rceil }{n}; \left\{L_i\right\} ^n_{i=1} \right) \right \} 
\end{align}

This interval can be interpreted as any arbitrary value of $y \in \mathcal{Y}$ for the specific input point $X_{n+1}$, for which the nonconforming score $L(X_{n+1},y)$ is below the adjusted empirical quantile of the nonconformal score computed on the available dataset $\mathcal{D}_n$. We can observe the direct equivalence to the rank statistics with the difference being the computation of the quantile from $\left\{L_i\right\}^n_{i=1}$ instead of $\left\{Y_i\right\}^n_{i=1}$. This is a very general definition that can be used to derive different CP methods. Most of the CP methods are distinguished based on two aspects: (1) the selected nonconformal score; (2) the method to guarantee the exchangeability of $\left \{\left(X_i,Y_i\right)\right\}^{n+1}_{i=1}$. 
% \begin{enumerate}
% 	\item The selected nonconformal score. Various scores can be defined with the property of smaller values indicating more conformal values between predicted and true values. 
% 	\item The method to guarantee the exchangeability of $(X_i,Y_i)^{n+1}_{i=1}$. This means that the underlying distribution does not depend on the ordering of the values or that it should be equivalent to add $(X_{n+1}, Y_{n+1})$ into the available dataset and predict on a different point from $(X_i, Y_i)$. For example, if the model $\hat{f}$ is trained on the same dataset $D_n$, then the exchangeability is violated since the nonconformal score will be computed on data points seen by the model, while the prediction would be performed on an unseen data point. Various techniques exist to address this issue. The most straightforward approach is called \textit{Split CP} and consist in training the ML model on a different dataset $D_m$ that is generated from the same underlying distribution of $D_n$. The nonconformal score is then computed on $D_n$ which is data not seen by the model as it is the case for $(X_{n+1}, Y_{n+1})$. Another approach that does not require two different datasets is the \textit{Full CP}, which guarantees the exchangeability by requiring retraining of the model for each prediction. In practice, this approach is very computational expensive and thus not widely used.
% \end{enumerate}

Combining the most natural choices for addressing these two aspects leads to the \textit{Split CP} method with absolute residuals. Two different datasets are used: (1) $\mathcal{D}_m=\left\{X_j,Y_j\right\}^m_{j=1}$ for training/fitting the ML model $\hat{f}_m$; and (2) $\mathcal{D}_n=\left\{X_i,Y_i\right\}^n_{i=1}$ to compute the residuals $L_i = |Y_i - \hat{f}_m(X_i)|$, and thus obtain the corresponding adjusted quantile $\hat{q}_{n, \alpha}=\hat{q}_n \left( \frac{\lceil (1-\alpha)(n+1) \rceil }{n}; \left\{L_i \right\}^n_{i=1} \right)=L^r_u$. $L^r_u$, as for $Y^r_u$ in the rank statistics, corresponds to the $u$\textsuperscript{th} value in the ranked set of residuals $\left\{L^r_i\right\}^n_{i=1}$ that will guarantee the desired confidence level. The resulting prediction interval for a new data point $\left\{X_{n+1},Y_{n+1}\right\}$ is:
\begin{align}
	%\mathbb{P} &\left ( Y_{n+1} \in [\hat{f}(X_{n+1})-\hat{q}_{n,\alpha}, \hat{f}(X_{n+1})+\hat{q}_{n,\alpha}]\right ) \geq 1 - \alpha \\ 
	%& \quad \quad \quad \quad \quad \quad \quad \quad \mathrm{or} \nonumber \\
	\mathbb{P} &\left ( Y_{n+1} \in [\hat{Y}_{n+1}-L^r_u, \hat{Y}_{n+1}+L^r_u]\right ) \geq 1 - \alpha.
\end{align}

As mentioned earlier, such an approach is very general and applies to any ML model. However, it has an important weakness due to the fact that it assigns the same interval length $\pm L^r_u$ to all predictions. This means that the prediction interval cannot adapt to local regions where the model might be more or less accurate. It is desirable that the regions of the input space $\mathcal{X}$ where the ML model is more accurate to have less uncertainty than the regions of $\mathcal{X}$ where the model is less accurate. This property is called local adaptivity and there are CP approaches that try to address it. One of the main approaches is called \textit{Studentized Residuals CP} (SRCP) and uses a score function that normalizes the residuals using an estimate of their standard deviation. This estimate is a second trained model $\hat{\sigma}_m$ on the dataset $\mathcal{D}_m=\left \{ X_j,\left|Y_j-\hat{f}_m(X_j) \right| \right \}^m_{j=1}$ and the resulting score is expressed as:
\begin{align}
	L_i = \frac{\left| Y_i-\hat{f}_m(X_i) \right|}{\hat{\sigma}_m(X_i)}, \quad i=1,2,\ldots,n.
\end{align}

The corresponding adjusted residual is obtained $\hat{q}_{n,\alpha}=L^r_u$ for the desired confidence level in the same way as the previous examples. The resulting prediction interval for a new data point $\left\{X_{n+1},Y_{n+1}\right\}$ becomes:
\begin{align}
	%\mathbb{P} &\left ( Y_{n+1} \in [\hat{f}(X_{n+1})-\hat{\sigma}_m(X_{n+1})\hat{q}_{n,\alpha}, \hat{f}(X_{n+1})+\hat{\sigma}_m(X_{n+1})\hat{q}_{n,\alpha}]\right ) \geq 1 - \alpha \\ 
	%& \quad \quad \quad \quad \quad \quad \quad \quad \mathrm{or} \nonumber \\
	\mathbb{P} &\left ( Y_{n+1} \in [\hat{Y}_{n+1}-\hat{\sigma}_m(X_{n+1})L^r_u, \; \hat{Y}_{n+1}+\hat{\sigma}_m(X_{n+1})L^r_u]\right ) \geq 1 - \alpha.
\end{align} 

% One of the drawbacks of this method is that it adapts the prediction intervals only based on the estimate of the local standard deviation, which in some cases can be misleading. The \textit{Quantile Regression CP} addresses these drawbacks by using again two ML models trained on $D_m=(X_j,Y_j)^m_{j=1}$, but that instead of predicting the best estimate value and its standard deviation, they predict directly the desired quantiles for the selected confidence level $\alpha$. The two models are $\hat{f}^{\alpha/2}_m$, and $\hat{f}^{1-\alpha/2}_m$ that predict the lower and upper quantile, respectively. These models are trained using quantile regression techniques. The score for this CP approach is also computed on the separate dataset $D_n$ and is expressed as:
% \begin{align}
% 	L_i = \max \left \{ \hat{f}^{\alpha/2}_m(X_i) - Y_i, Y_i - \hat{f}^{1-\alpha/2}_m(X_i) \right \}, \quad i \in D_n
% \end{align}

% The corresponding adjusted residual is obtained $\hat{q}_{n,\alpha}=L^r_u$ for the desired confidence levels and the resulting prediction interval for a new data point $(X_{n+1},Y_{n+1})$ becomes:

% \begin{align}
% 	\mathbb{P} &\left ( Y_{n+1} \in [\hat{f}^{\alpha/2}_m(X_{n+1})-\hat{q}_{n,\alpha}, \hat{f}^{1-\alpha/2}_m(X_{n+1})-\hat{q}_{n,\alpha}]\right ) \geq 1 - \alpha
% \end{align} 

%%%%%%%%%%%%%%%%%%%%%%%%%%%%%%%%%%%%%%%%%%%%%%%%%%%%%%%%%%%%%%%%%%%%%%%%%%%%%%%%
\section{Demonstration Examples}
\label{section:demonstration}
%%%%%%%%%%%%%%%%%%%%%%%%%%%%%%%%%%%%%%%%%%%%%%%%%%%%%%%%%%%%%%%%%%%%%%%%%%%%%%%%

The presented methods for UQ of ML models in Section \ref{section:methodologies} will provide different uncertainty predictions that are not straightforward to assess and evaluate. We define the following desired properties for the predicted uncertainties:
\begin{itemize}
    \item \textit{Data coverage with a pre-defined confidence}. For example, a method can predict uncertainty bands that include $95 \%$ of the data points.
    \item \textit{Minimize coverage that is needed to ensure the pre-defined confidence}. For example, excessively large uncertainty bands can always be found that will cover all the data points. However, this obviously is not meaningful. 
    \item \textit{Extrapolation regions should be associated with large uncertainties}. Due to the lack of data in these regions, UQ methods should take this into account by predicting a larger uncertainty as an indication that one should have less confidence in the ML predictions in this region.
\end{itemize}

These properties for each UQ method are tested on two demonstration examples. The first one is a simple analytical example that will allow one to better understand the performance of each method against a known solution. The second is a realistic nuclear engineering application using neutron flux measurement data from the SAFARI-1 reactor.

%%%%%%%%%%%%%%%%%%%%%%%%%%%%%%%%%%%%%%%%%%%%
\subsection{Analytical GP}
\label{section:demonstration-simple}
%%%%%%%%%%%%%%%%%%%%%%%%%%%%%%%%%%%%%%%%%%%%

In this example, the data are generated from a known GP and consequentially the analytical confidence intervals are available. The GP is defined as $\mathcal{GP}\left( \mu (x), k(x,x') \right): \mathcal{X} \rightarrow \mathbb{R}$, where $\mathcal{X}$ is the region $[0,10]$, $\mu (x)$ is the mean function, and $k(x,x')$ is the covariance function. The following mean function is used in the GP model:
\begin{equation}
    \mu (x) = x + 0.02 x^2 + 5 \sin{x}
\end{equation}

\begin{figure}[!ht]
	\centering
\includegraphics[scale=0.4]{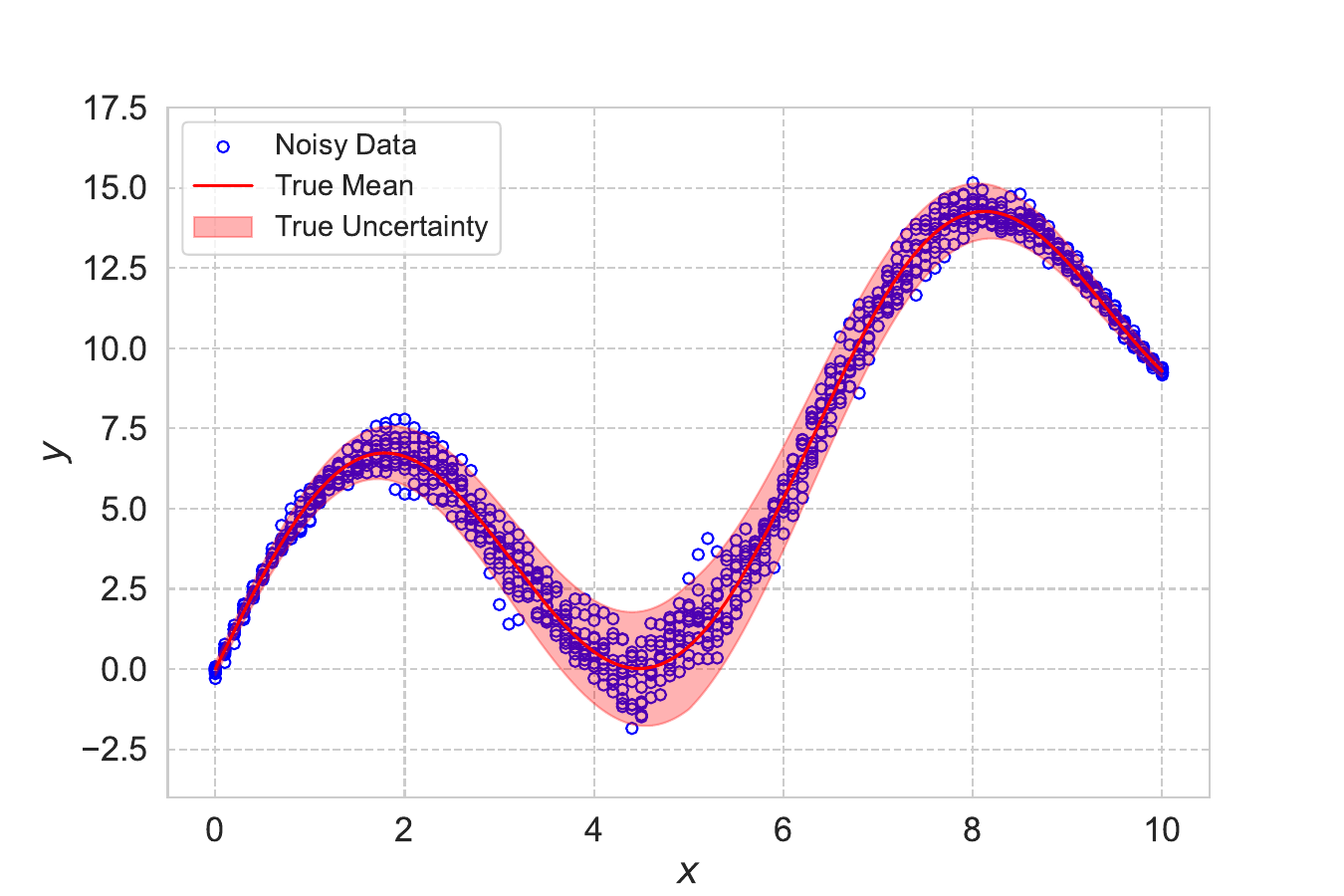}
	\caption{Generated dataset from GP. 
    The ``True uncertainty'' means $95 \%$ confidence interval from the GP prediction.}
	\label{figure:gp_dataset}
\end{figure}

The Mat{\'e}rn 5/2 covariance kernel is selected with a length parameter of $0.2$. In order to investigate regions of varying uncertainty, the standard deviation of the GP increases linearly from the minimum value of $0.1$ at $x=0$, to the maximum value of $1.0$ at $x=5$, and then decreases linearly back to the minimum value of $0.1$ at $x=10$. From this analytical model, $10$ GP realizations are sampled at $100$ equidistant points. This creates a total dataset of $1000$ data points visualized in Figure~\ref{figure:gp_dataset}. 
This dataset is used to train different ML models with various UQ methods for a comparison. It is important to note that the 95\% confidence interval is used as the desired the uncertainty bounds. This interval is shown in Figure \ref{figure:gp_dataset} for the analytical model and is computed as $\mu(x) \pm 1.96 \, \sigma(x)$, where $\sigma(x)$ is the GP standard deviation at every $x$.

\begin{figure}[!ht]
	\centering
	\begin{subfigure}[b]{0.495\textwidth}
		\includegraphics[width=.99\textwidth]{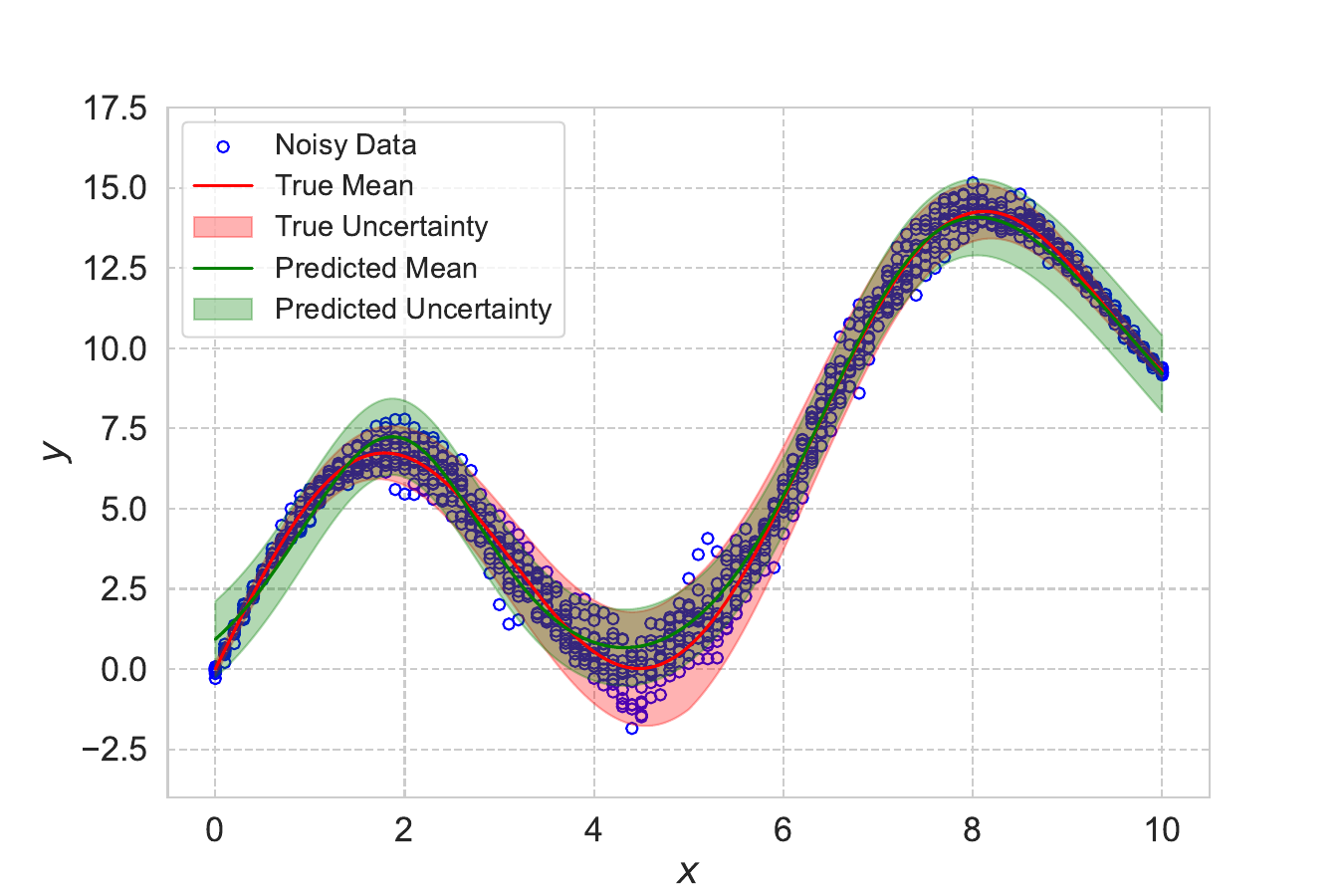}
		\caption{Split CP results for the DNN model. \vspace{10pt}}
		\label{figure:GP_SCP_DNN-a}
	\end{subfigure}
	\begin{subfigure}[b]{0.495\textwidth}
		\includegraphics[width=.99\textwidth]{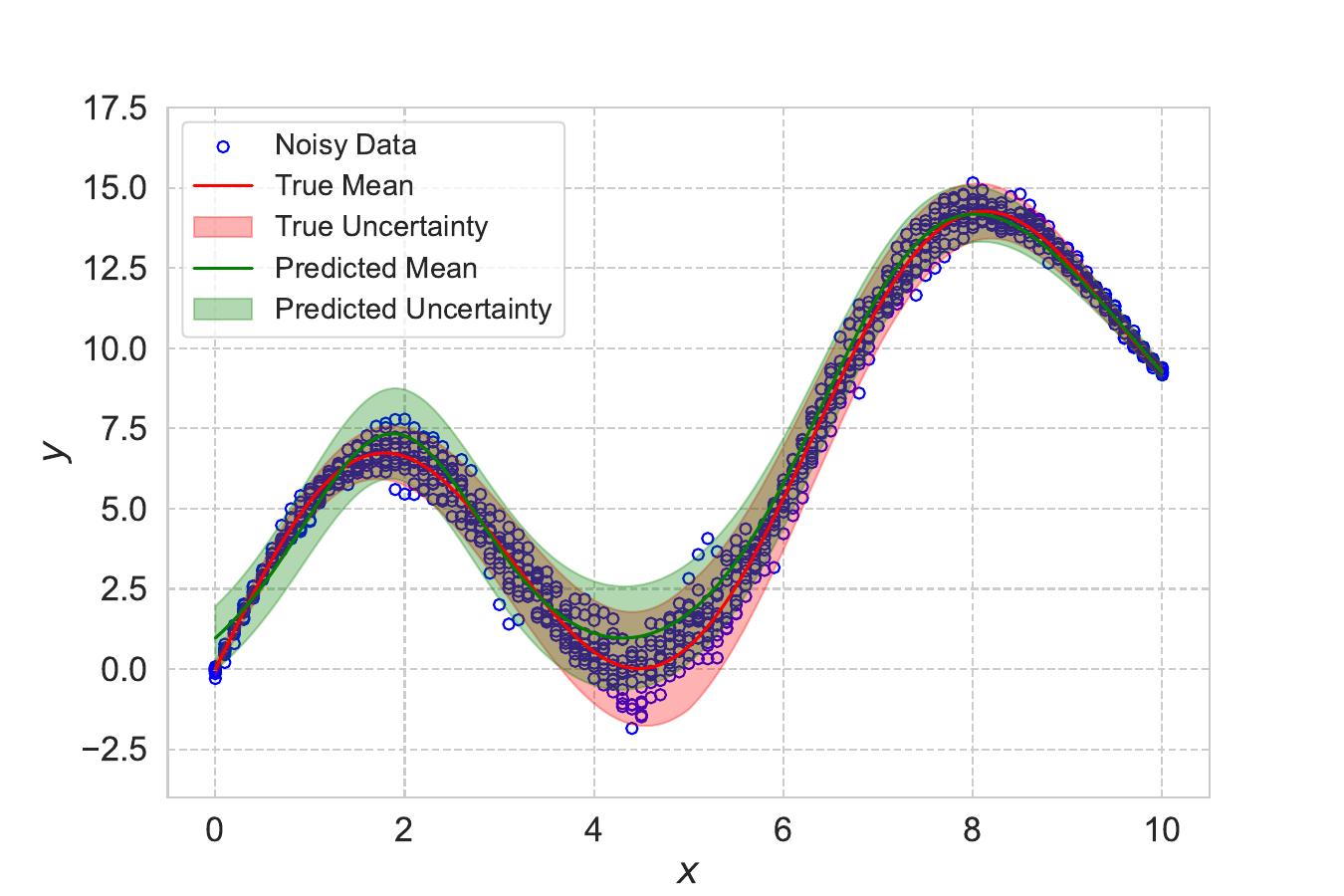}
		\caption{SRCP results for DNN models for both mean and residual models.}
		\label{figure:GP_SRCP_DNN-b}
	\end{subfigure}
        \\
	\begin{subfigure}[b]{0.495\textwidth}
		\includegraphics[width=.99\textwidth]{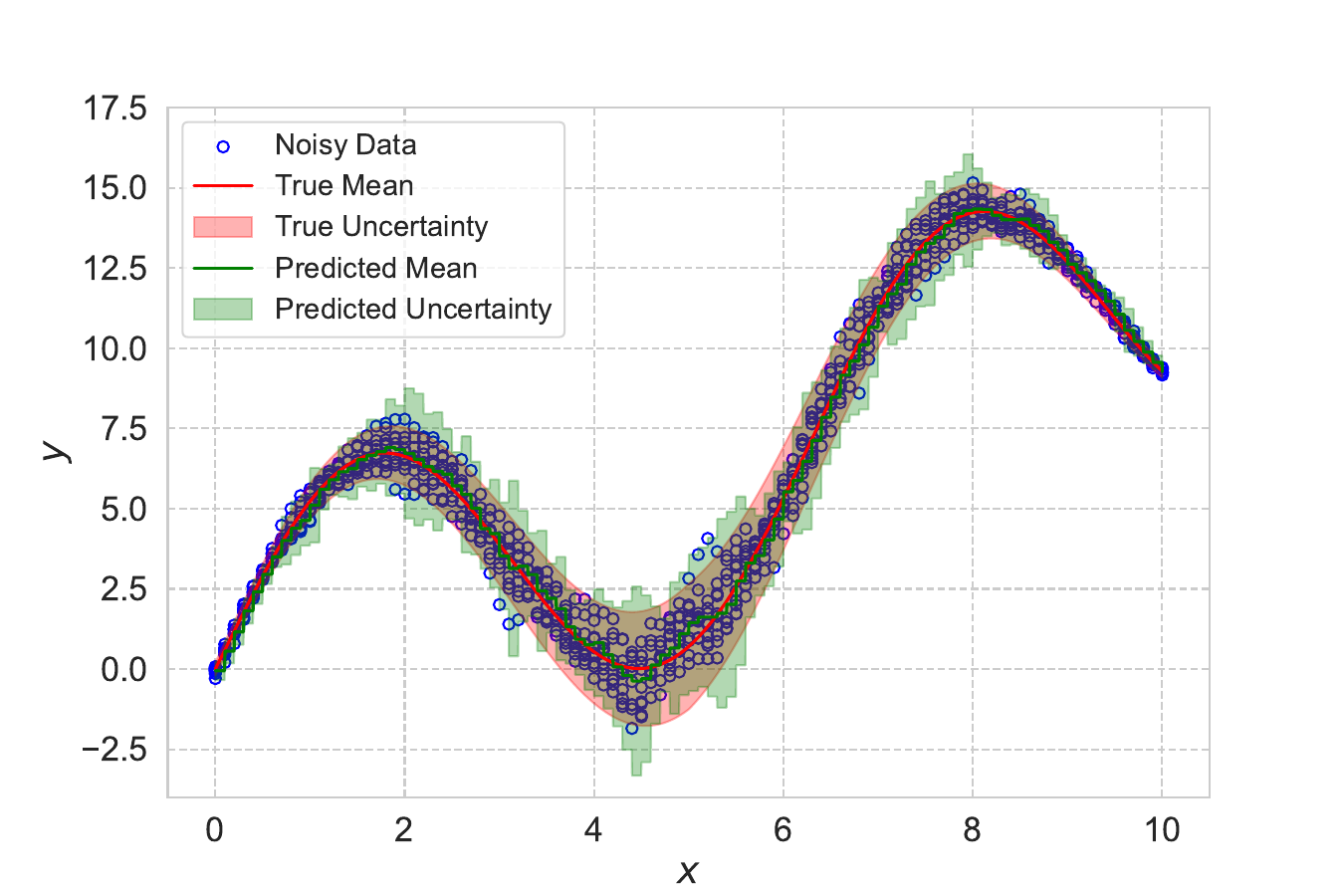}
		\caption{SRCP results for XGBoost models for both mean and residual models.}
		\label{figure:GP_SRCP_XGB-c}
	\end{subfigure}
	\begin{subfigure}[b]{0.495\textwidth}
	\includegraphics[width=.99\textwidth]{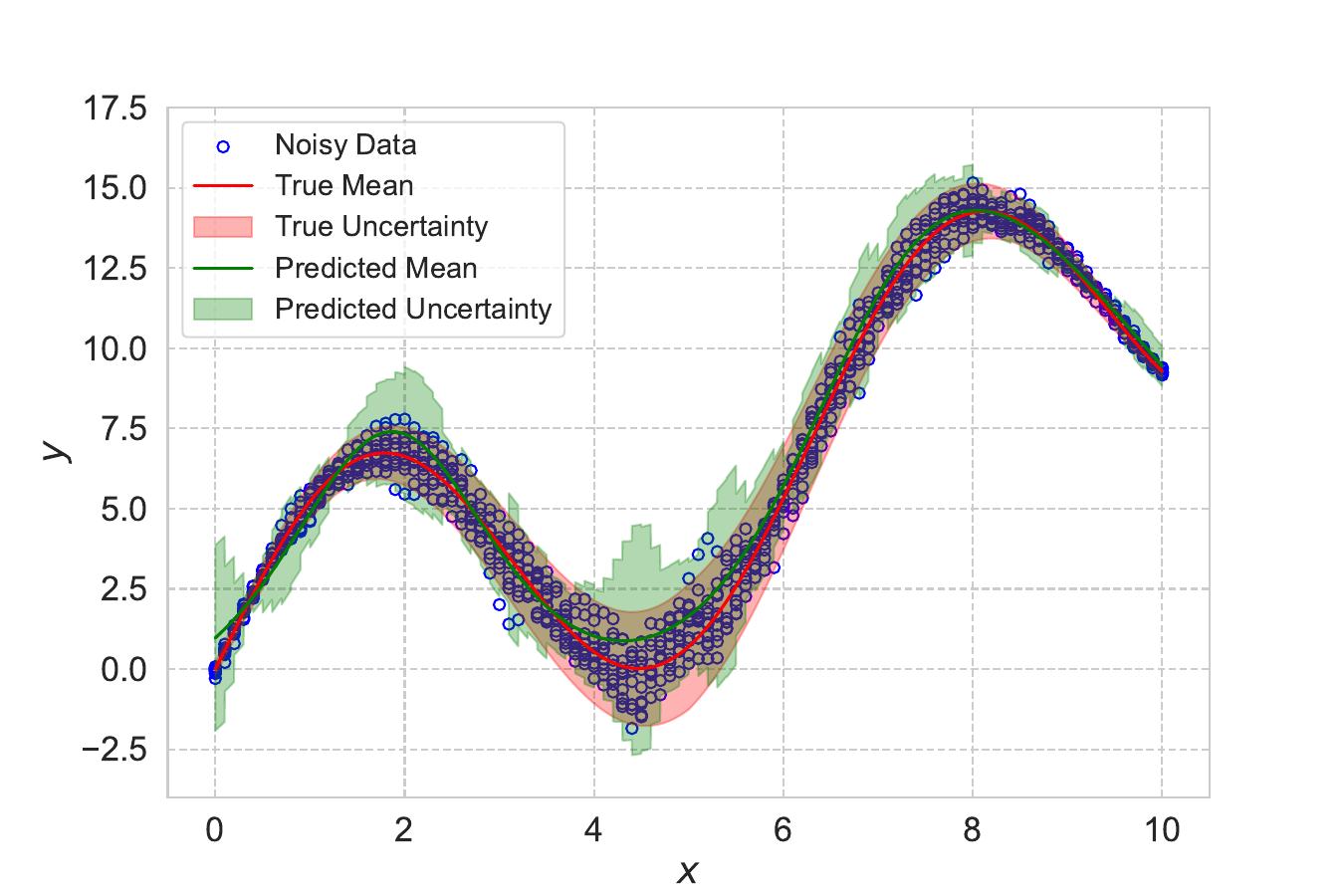}
		\caption{SRCP results for DNN model for mean and XGBoost for residual prediction.}
		\label{figure:GP_SRCP_DNN_XGB-d}
	\end{subfigure}
	\caption{UQ results for the analytical GP example using different CP methods on DNN and XGBoost models.}
	\label{figure:GP_CP}
\end{figure}

The results of the CP method applied to DNN and XGBoost models are shown in Figure~\ref{figure:GP_CP}. All the DNN models have the same architecture and hyperparameters. Four hidden layers are used with the following number of neurons per layer: $[200,500,500,200]$. The tanh activation function is used in each neuron. The L$_2$ regularization is used to avoid overfitting, while the Adam optimizer was employed for training. For XGBoost, a total of $200$ trees are used with a maximum depth of $12$. 

The results of the split CP method for the DNN are shown in Figure~\ref{figure:GP_SCP_DNN-a}. A good coverage of the data points is observe, but the method, as expected, does not adapt locally in the regions where the uncertainty is smaller or larger. The split CP method produces a constant uncertainty that is too large in the bounds of the domain and too small in the center where the uncertainty is larger. The SRCP method results using two DNNs, one for the mean prediction and one for the residual, are shown in Figure~\ref{figure:GP_SRCP_DNN-b}. The uncertainty now adapts better to the local regions, especially for $x > 7$. However, the results are still not very satisfactory in the remaining domain. This is primarily attributed to the residual DNN, which cannot predict the local uncertainties very accurately. This is even more evident when the XGBoost model is used for both mean and residual predictions. These models predict more accurately the residuals and their results presented in Figure~\ref{figure:GP_SRCP_XGB-c} show a good agreement with the analytical solution. In Figure~\ref{figure:GP_SRCP_DNN_XGB-d}, the results for SRCP for hybrid models using DNN for the mean prediction and XGBoost for the residual shows a slightly worse performance than the XGBoost results. Still these results consist an improvement over the DNN SRCP results.

\begin{figure}[!ht]
	\centering
	\begin{subfigure}[b]{0.495\textwidth}
		\includegraphics[width=.99\textwidth]{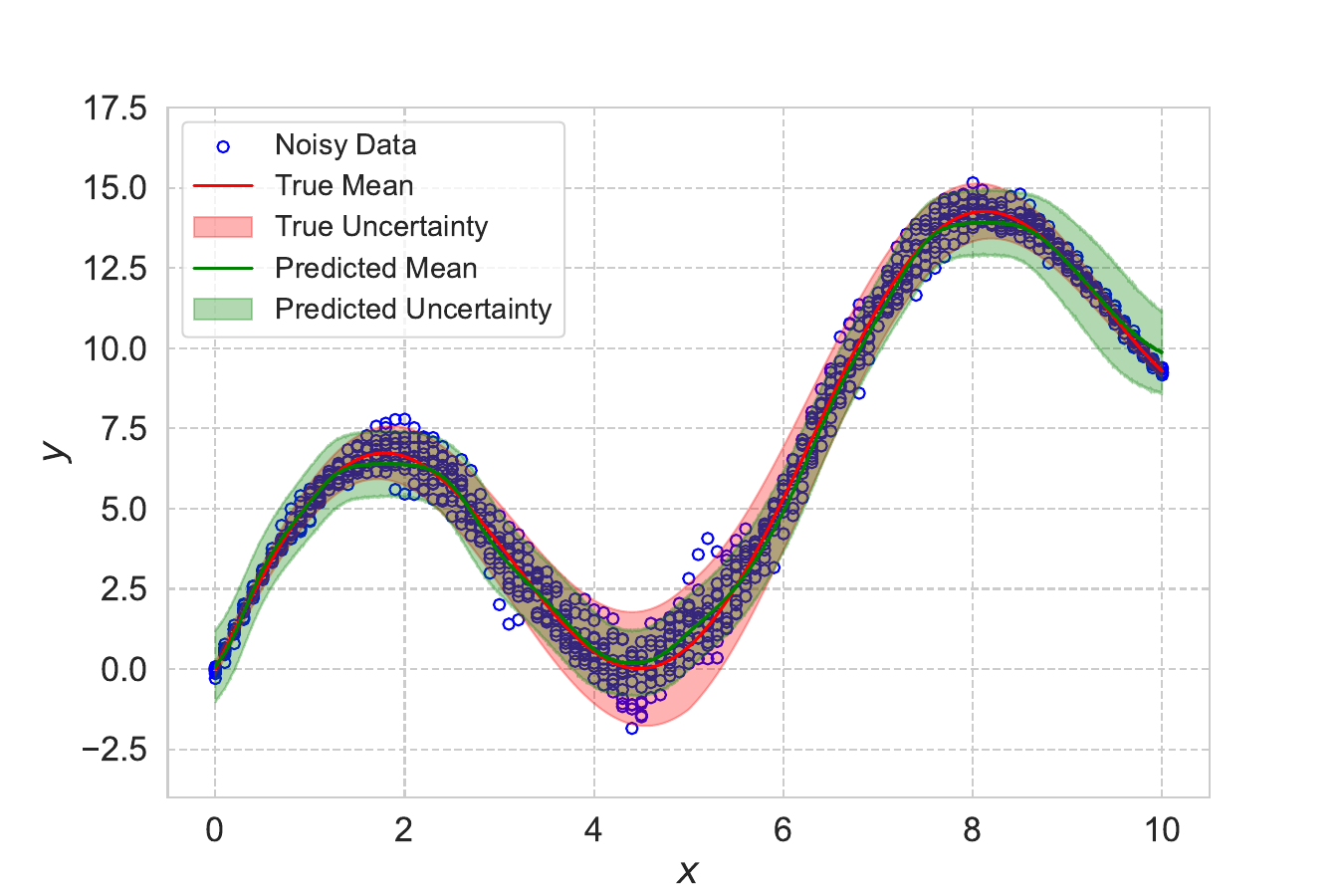}
		\caption{MCD}
		\label{figure:GP_MCD-a}
	\end{subfigure}
	\begin{subfigure}[b]{0.495\textwidth}
		\includegraphics[width=.99\textwidth]{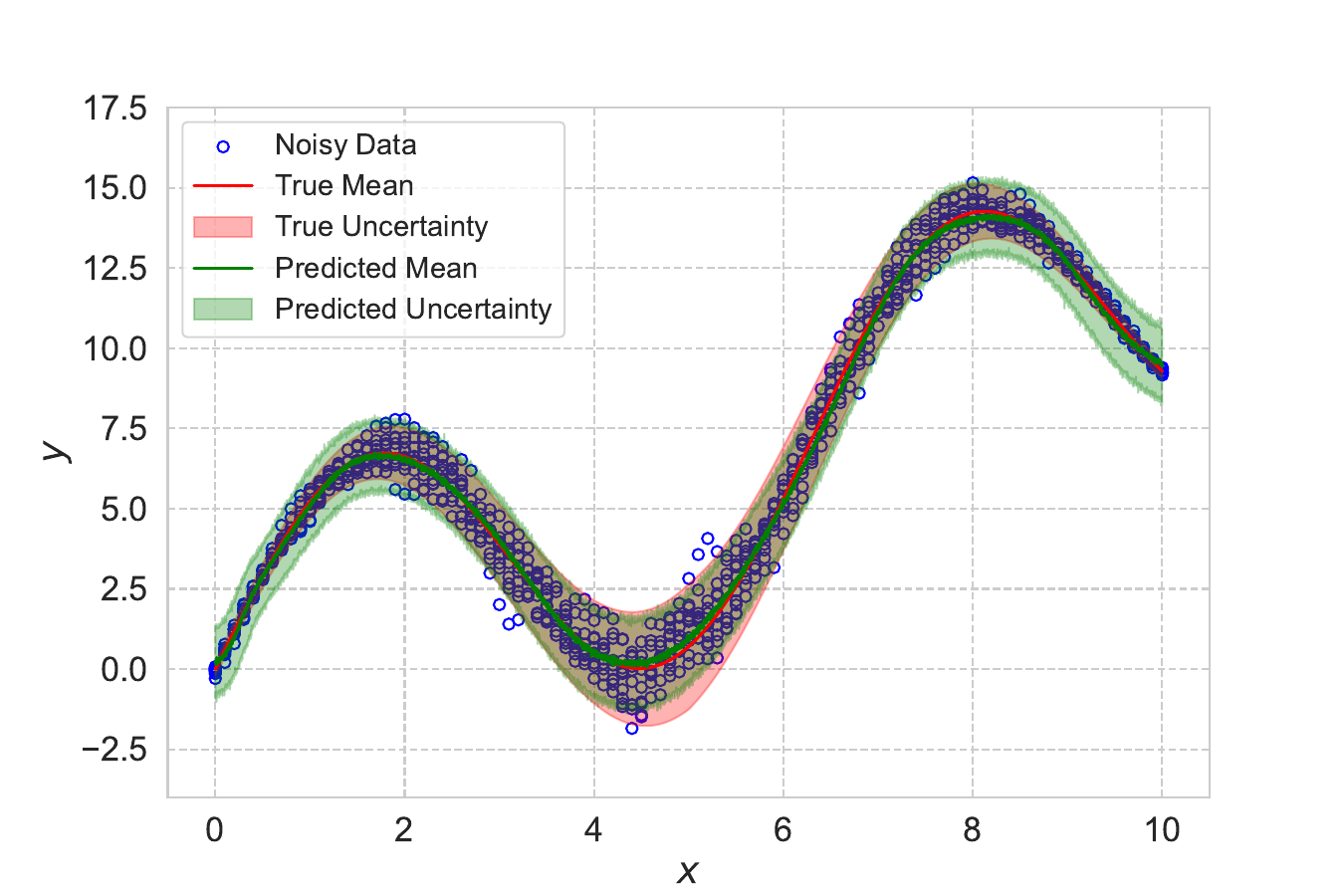}
		\caption{BNN}
		\label{figure:GP_BNN-b}
	\end{subfigure}
        \\
	\begin{subfigure}[b]{0.495\textwidth}
		\includegraphics[width=.99\textwidth]{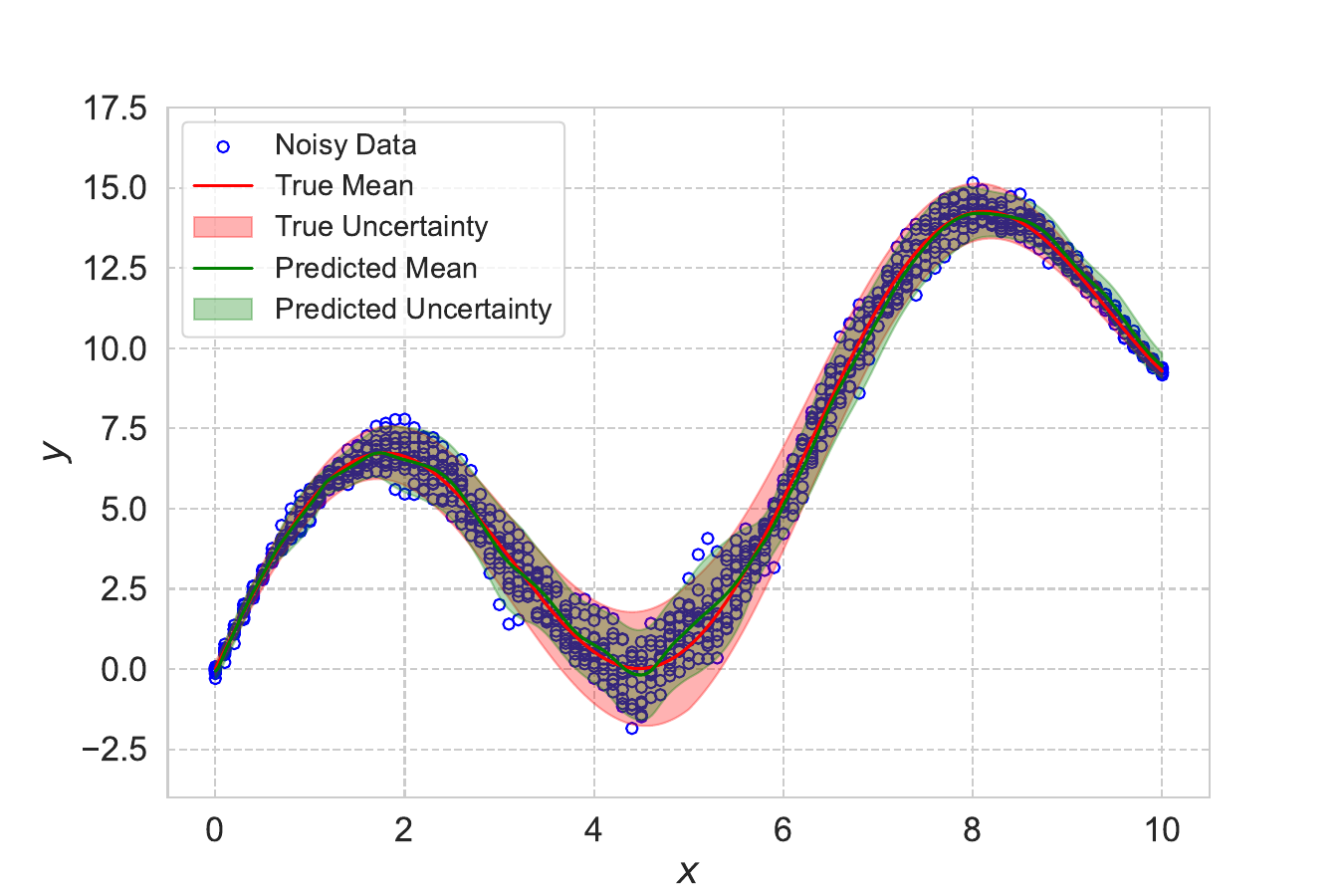}
		\caption{DE}
		\label{figure:GP_DE-c}
	\end{subfigure}
	\begin{subfigure}[b]{0.495\textwidth}
	\includegraphics[width=.99\textwidth]{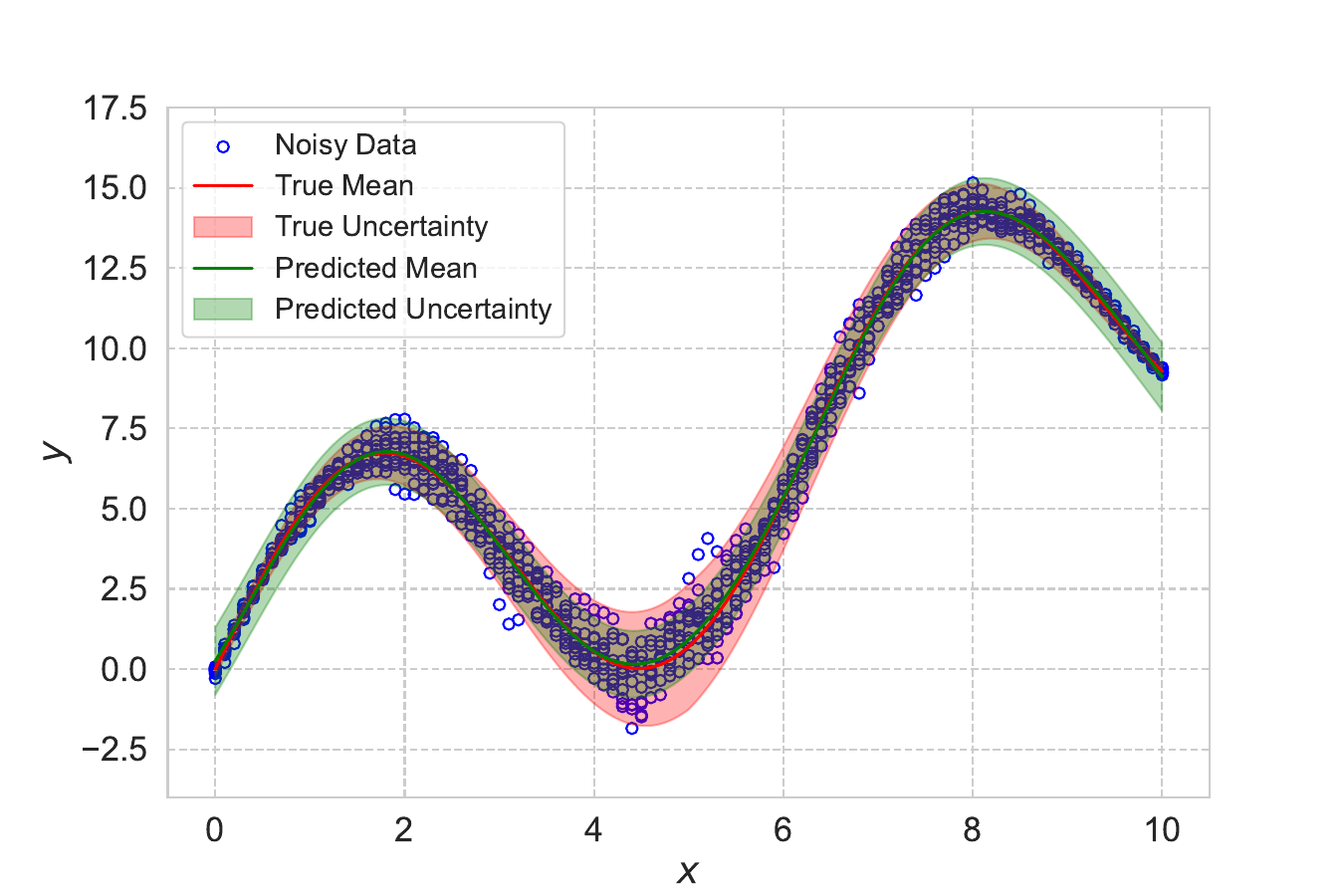}
		\caption{GP}
		\label{figure:GP_GP-d}
	\end{subfigure}
	\caption{Results for other UQ methods presented in Section \ref{section:methodologies}.}
	\label{figure:GP_Other}
\end{figure}

The results for the other UQ methods discussed in Section \ref{section:methodologies} are gathered in Figure~\ref{figure:GP_Other}. In Figure~\ref{figure:GP_MCD-a}, the MCD method shows a similar behavior to the split CP. The uncertainty prediction does not adapt locally with a rather constant value across the whole domain. It is important to note that the same DNN architecture is used with the CP analysis above, but with a dropout of $25 \%$. In Figure~\ref{figure:GP_BNN-b}, the BNN results cover more data points than the MCD and show better performance when compared against the analytical solution. The BNN also show some slight local adaptivity in the uncertainty prediction. The BNN architecture is more simplified than the DNN model, with only three hidden layers and 10 neurons in each layer. Such simpler architecture is an empirical choice based on our previous study \cite{yaseen2023quantification}. In Figure~\ref{figure:GP_DE-c}, the DE show impressive local adaptivity performance across the whole domain with results that are comparable to the SRCP with XGBoost. In Figure~\ref{figure:GP_GP-d}, the last method tested is a GP model, using different kernel that the one used to generate the data and by assuming homoscedastic uncertainty across the domain. The results are similar to the split CP with a constant uncertainty, as expected.

Overall, we observe that MCD, GP, and split CP perform similarly and predict a rather constant uncertainty across the whole domain. BNN provides a slight improvement in terms of local adaptivity and is the method that provides the largest data point coverage. Finally, DE and SRCP with XGBoost are the two methods that perform the best with their predicted uncertainty bounds being the closest to the analytical ones.

%%%%%%%%%%%%%%%%%%%%%%%%%%%%%%%%%%%%%%%%%%%%
\subsection{SAFARI-1 axial neutron flux measurements}
\label{section:demonstration-SAFARI-1}
%%%%%%%%%%%%%%%%%%%%%%%%%%%%%%%%%%%%%%%%%%%%

In this section, we will provide a demonstration of the UQ of ML models using some of the methods discussed above in a realistic nuclear engineering problem. In our previous work \cite{moloko2023prediction,moloko2024clustering}, ML models have been developed for predicting the assembly-wise axial neutron flux profiles in the SAFARI-1 research reactor, trained by measured data from historical operational cycles. These measurements entail irradiation of copper wire along the active length of the fuel-containing assemblies in the SAFARI-1 reactor core. ML models were trained for each fuel-containing assembly separately and prediction uncertainties were quantified. 
Each training data point consisted of a control bank position and an axial location of the measurement point, both treated as inputs, as well as a measured detector count value treated as the output. 
%A total of 13,497 data points were used for training the neural networks. 
The input and output features were standardized using z-score normalization, ensuring a mean of zero and a standard deviation of one. To address time-based measurement variations, pre-processing included aligning scanned data to a common reference time via exponential decay corrections, as detailed in \cite{moloko2023prediction,moloko2024clustering}.

% Examples, which will be used for illustration in this section, are a combination of new and old (but unpublished) results. Thus, in the case of DNN models, the DE model was added; for all DNNs, a $k$-fold cross-validation was introduced, and a different tool was used for hyperparameter tuning; this can be seen as evolution
% of the approach followed in \cite{moloko2023prediction,moloko2024clustering}. GP model illustrations are unpublished results originating from studies presented in the same references. 
%
In this study, hyperparameter optimization for neural network models of interest -- DNN with MCD, DE, and BNN -- was performed using Neural Network Intelligence (NNI) framework \cite{nni2021} with a random search. In all cases, the Adam optimization algorithm was employed, Negative Log-Likelihood (NLL) was used as a loss function, and Rectified Linear Unit (ReLU) was used as an activation function.
To enhance performance and robustness, $k$-fold cross-validation with $k=5$ was employed for both hyperparameter tuning and training. This method splits the dataset into $k$ non-overlapping folds, training on $(k-1)$ folds and validating on the remaining fold in $k$ iterations. This approach provides reliable estimates of mean error and variance while balancing computational efficiency and performance. 

For the sake of illustration, the fuel assemblies in SAFARI-1 core positions D5 (towards core center) and H6 (on the core periphery) were selected for demonstration with MCD, DE, BNN and GP as the UQ methods. The two outputs of the models are the predicted mean, $\mu$, and the uncertainty (standard deviation), $\sigma$. As an example, the optimal architecture and hyperparameter for D5 assembly based on these UQ models, are presented in Table~\ref{tab:hyperparameters}. Furthermore, positions C7 and E5 hosting control assemblies were selected for comparison of flux profile prediction and UQ by GP and MCD. Other fuel and control assemblies show similar behavior in the UQ analysis, therefore they will not be presented in this paper.

\begin{table}[htb!]
  \centering\footnotesize
  \caption{Hyperparameters for MCD, DE, and BNN for fuel assembly D5.}
  \label{tab:hyperparameters}
\begin{tabular}{lccc}
\toprule 
Hyperparameters  & MCD & DE  & BNN\tabularnewline
\midrule 
Number of inputs  & 2  & 2  & 2\tabularnewline
Number of outputs  & 2\textsuperscript{**}  & 2\textsuperscript{**} & 2\textsuperscript{**}\tabularnewline
Number of layers  & 3  & 3  & 3\tabularnewline
Nodes per layer  & (247, 322, 486)  & (141, 434, 244)  & (323, 81, 311)\tabularnewline
Activation function  & ReLU  & ReLU  & ReLU\tabularnewline
Learning rate  & \num{1.9E-03}  & \num{3.2E-03}  & \num{1.9E-03} \tabularnewline
Loss function  & NLL  & NLL  & NLL\tabularnewline
Optimizer  & Adam  & Adam  & Adam\tabularnewline
Batch size  & 128  & 128  & 64\tabularnewline
Number of epochs  & 1000  & 1000  & 1000\tabularnewline
Dropout rate  & 0.11 & --  & -- \tabularnewline
% Training samples  & 11047  & 11047  & \added{40}\tabularnewline
% Test samples  & 684  & 684  & \added{7}\tabularnewline
% Validation samples  & 1949  & 1949  & --\tabularnewline
\midrule 
% \multicolumn{4}{l}{\added{\textsuperscript{*}BNN VI uses the same architecture and hyperparameters
% as those of the Keras DNN models}}\tabularnewline
\multicolumn{4}{l}{\textsuperscript{**}mean and uncertainty}\tabularnewline
\end{tabular}  
\end{table}    

The SAFARI-1 problem virtually contains all five sources of uncertainty presented in Figure~\ref{figure:uncertainty-source-data-driven-ML}. The \textit{Data Noise} uncertainty originates from the statistical error of the order of 7\%, contained in the copper wire measurements. Additional measurements uncertainties could arise from several factors, such as the imprecise positioning of the reactor control rods during the experiment, fluctuations in reactor power, incorrect placement of the irradiated copper wire, etc. These types of uncertainty have been discussed and quantified in \citep{macconnachie2019quantification,macconnachie2021measurement} for the McMaster Nuclear Reactor, for example. The \textit{Data Coverage} and \textit{Extrapolation} uncertainties are present because the control bank positions from the historical cycles cover unevenly only roughly 100~mm from the total range of 600~mm. Finally, the \textit{Imperfect Model} and \textit{Training} uncertainties would also be present if insufficient care were taken in the process of ML model design and training.

\begin{figure}[!ht]
	\centering
	\begin{subfigure}[b]{0.495\textwidth}
		\includegraphics[width=.99\textwidth]{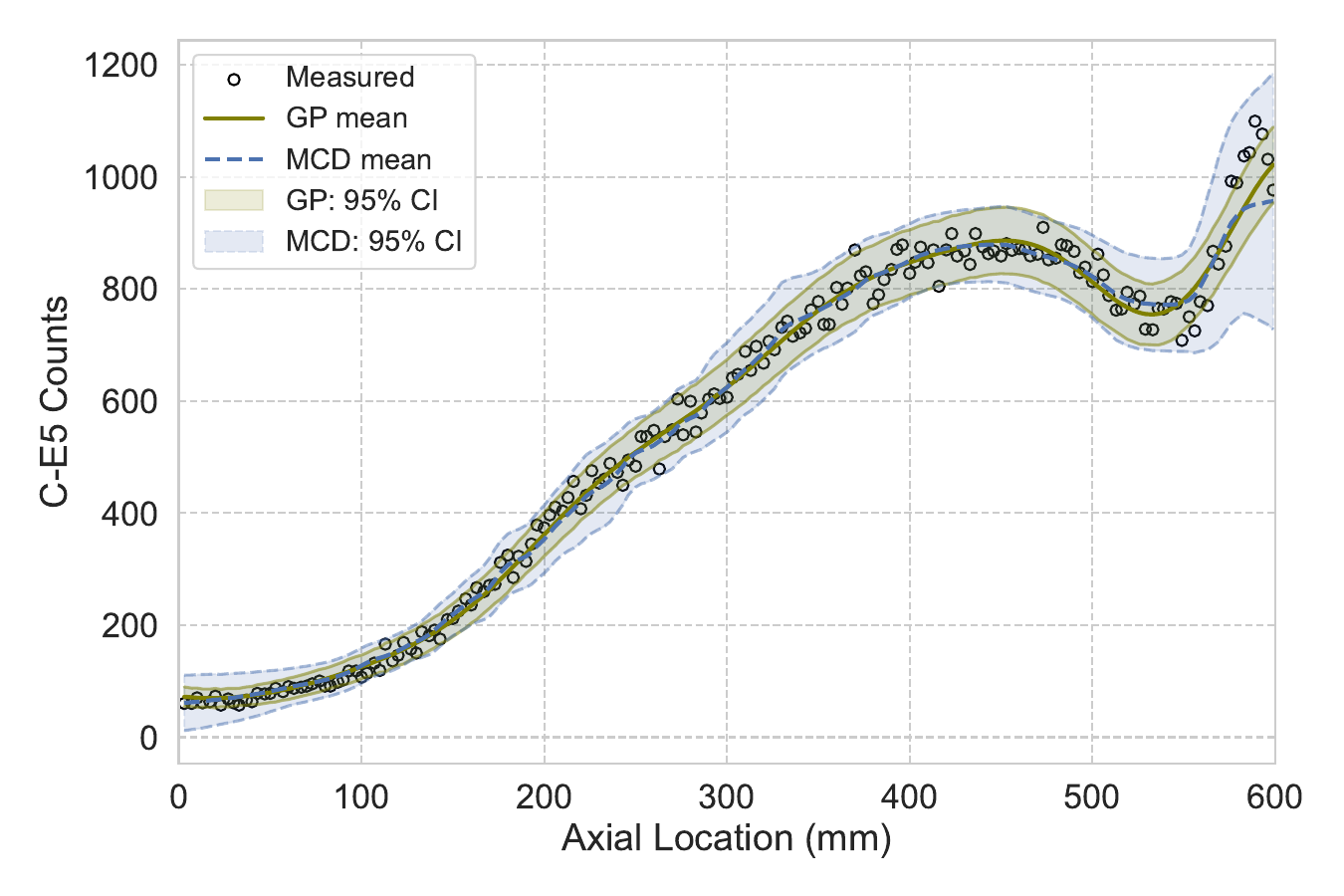}
		\caption{}
		\label{figure:SAFARI-1-Prediction-and-UQ-a}
	\end{subfigure}
	\begin{subfigure}[b]{0.495\textwidth}
		\includegraphics[width=.99\textwidth]{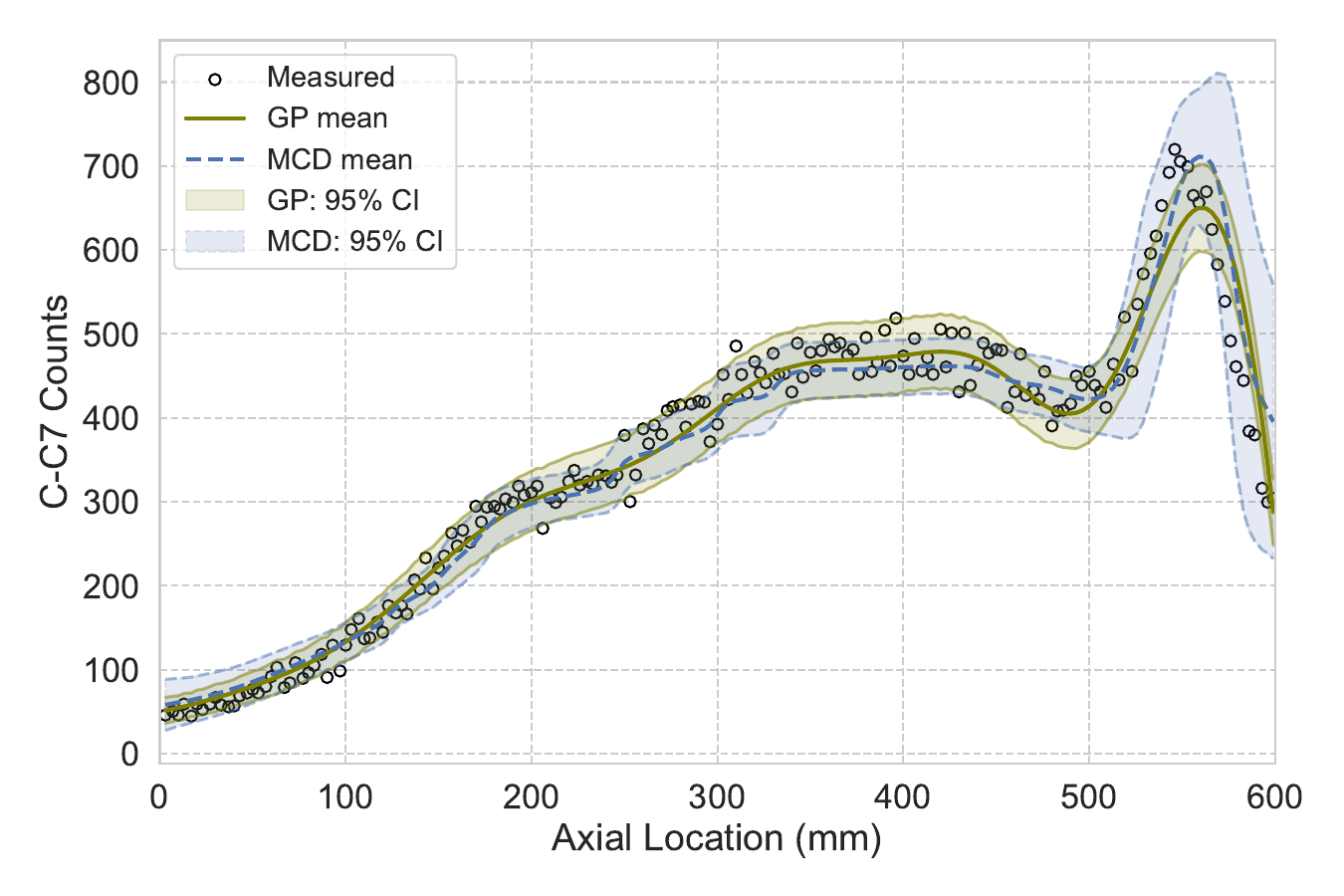}
		\caption{}
		\label{figure:SAFARI-1-Prediction-and-UQ-b}
	\end{subfigure}
	\caption{Prediction and UQ of the axial flux distribution in SAFARI-1 control assemblies (left E5, right C7) by GP and MCD methods.}
	\label{figure:SAFARI-1-Prediction-and-UQ}
\end{figure}

\begin{figure}[!ht]
	\centering
	\begin{subfigure}[b]{0.495\textwidth}
		\includegraphics[width=.99\textwidth]{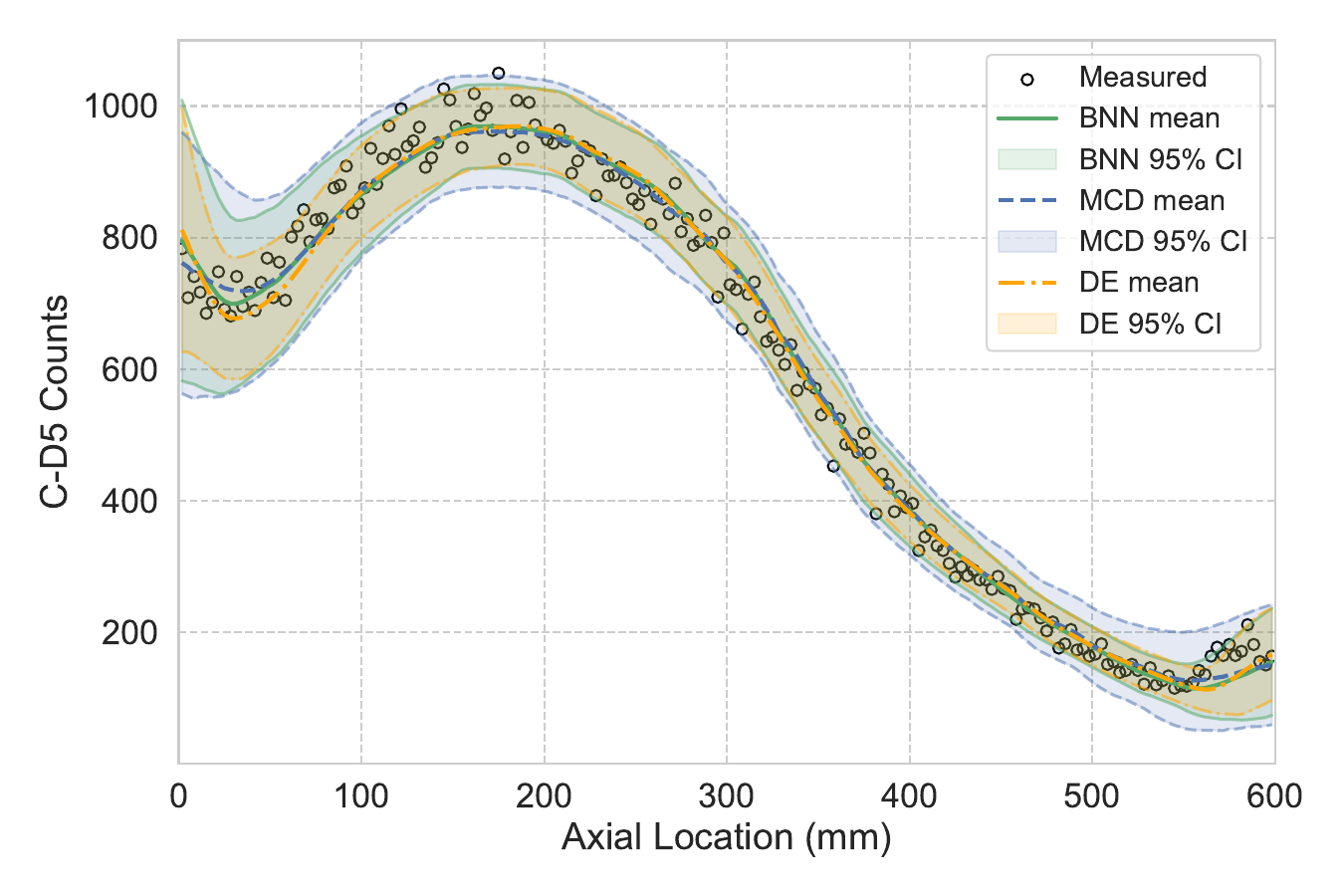}
		\caption{}
		\label{figure:SAFARI-1-Prediction-and-UQ-c}
	\end{subfigure}
	\begin{subfigure}[b]{0.495\textwidth}
		\includegraphics[width=.99\textwidth]{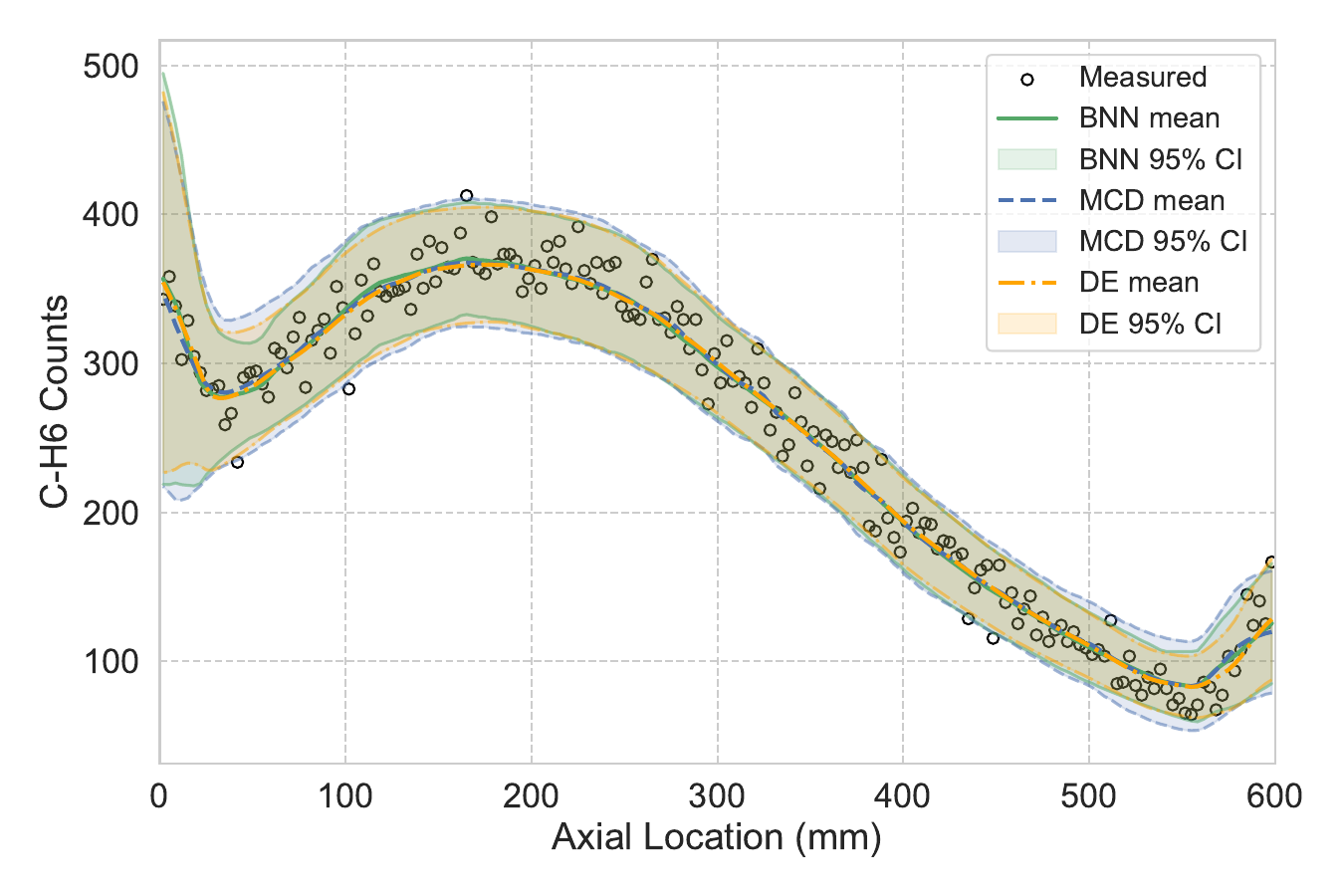}
		\caption{}
		\label{figure:SAFARI-1-Prediction-and-UQ-d}
	\end{subfigure}
	\caption{Prediction and UQ of the axial flux distribution in SAFARI-1 fuel assemblies (left D5, right H6) by BNN, MCD and DE methods.}
	\label{figure:SAFARI-1-Prediction-and-UQ-MCD-BNN-DE}
\end{figure}

Figure~\ref{figure:SAFARI-1-Prediction-and-UQ} illustrates the axial flux profile predicted by DNN and GP and associated uncertainty bands (corresponding to the 95\% confidence interval -- CI) in reactor core positions E5 and C7 (both containing control rods) for a cycle that is not used in training of models. As one may observe in Figure~\ref{figure:SAFARI-1-Prediction-and-UQ-a}, DNN MCD and GP yield consistent predictions, which indicates that the main source of uncertainty in this case is the \textit{Data Noise}. This is supported by the uncertainty band that tightly covers the experimental points (though it is slightly wider for DNN MCD). In Figure~\ref{figure:SAFARI-1-Prediction-and-UQ-b}, there is some observable discrepancy between predictions by two methods, which can be an indication of contribution by the \textit{Imperfect Model} and \textit{Training} uncertainties. Effect of the \textit{Data Coverage} and \textit{Extrapolation} uncertainties is small, otherwise they would also affect the result in Figure~\ref{figure:SAFARI-1-Prediction-and-UQ-a}.

The results for the fuel assemblies in core positions D5 and H6 are presented in Figure~\ref{figure:SAFARI-1-Prediction-and-UQ-MCD-BNN-DE}, along with their corresponding 95\% CIs. Consistent with the findings from Figure~\ref{figure:SAFARI-1-Prediction-and-UQ}, all three methods -- BNN, MCD, and DE -- demonstrate excellent predictive performance by closely aligning with the measured data. They accurately capture the overall trend, periodicity, and amplitude of the counts along the axial locations for both datasets. 
%That is for the smoother, less noisy data of D5, situated near the core center, and the more variable, noise-prone data of H6, which demonstrates heightened fluctuations owing to its peripheral position.
%
In terms of UQ, DE exhibited the narrowest 95\% CIs, reflecting high confidence in its predictions. BNN provided moderately wider CIs compared to DE, offering a balance between precision and robustness. MCD, on the other hand, displayed the widest CIs, ensuring the most conservative uncertainty estimates. These differences suggest that DE, while confident, may be less robust in regions of high variability, where it has the potential of underestimating uncertainty. Conversely, MCD is more robust in such regions due to its broader CIs, but this can lead to overestimation of uncertainty in stable regions. BNN seems to strikes a balance between these two approaches, balancing precision and adaptability. These observations hold consistently for both assembly, further validating the strengths and limitations of each method across different core positions.

%%%%%%%%%%%%%%%%%%%%%%%%%%%%%%%%%%%%%%%%%%%%%%%%%%%%%%%%%%%%%%%%%%%%%%%%%%%%%%%%
\section{Discussions on the need of ML VVUQ}
\label{section:discussions}
%%%%%%%%%%%%%%%%%%%%%%%%%%%%%%%%%%%%%%%%%%%%%%%%%%%%%%%%%%%%%%%%%%%%%%%%%%%%%%%%

For physics-based computational models, VVUQ has been very widely investigated and a lot of methodologies have been developed \cite{oberkampf2010verification}. In brief, \textit{verification} aims to identify, quantify, and reduce errors during the mapping from mathematical model to a computational model, \textit{validation} aims to determine the degree of accuracy of the model in representing real physics. Verification consists of two steps, \textit{code verification} to access the reliability of the software coding with numerical algorithm verification and software quality engineering, and \textit{solution/calculation verification} to evaluate the numerical accuracy of the solutions to a computer code. However, for ML models, there are very few quantitative metrics with which to determine the VVUQ performance. 
Some standard VVUQ activities for physical models can be followed without change for ML models. However, the data-driven nature of ML incurs the need for additional approaches to verify, validate, and evaluate these algorithms. Scientific datasets, either from experimentation, simulation or combination of the two, are often high-dimensional, noisy, heterogeneous, low-signal-to-noise, and multiscale. Therefore, a well-formulated and robust VVUQ framework for ML models is needed for nuclear engineering, with an emphasis on design and safety analysis of advanced nuclear reactors.

\begin{figure}[!htb]
	\centering
	\includegraphics[width=.95\linewidth]{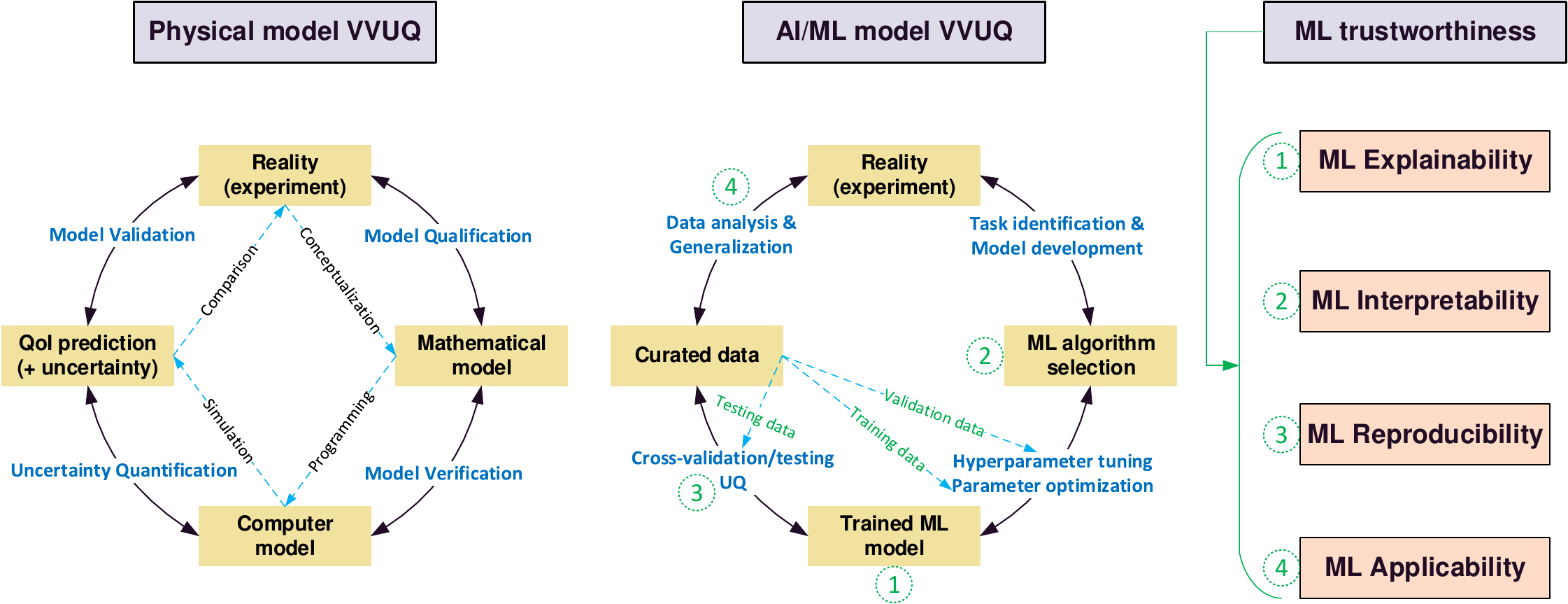}
	\caption{Illustration of the major components of the ML VVUQ framework, compared with the physical model VVUQ framework.}
	\label{figure:VVUQ}
\end{figure}

The VVUQ of ML models is very different from the VVUQ of physical models. Figure~\ref{figure:VVUQ} illustrates the major components of the ML VVUQ framework, with comparison to physical model VVUQ. Note that the ML VVUQ framework is only tentative; a more complete version with quantifiable metrics is under development and will be published in a future work. 
In ML, the computer model is inherent to experimental data, whereas in physical model development, the computer model is inherent to the mathematical model from conceptualization of the real phenomena. The ML model development steps involves feature selection, model architecture, loss/activate functions selection, etc. 
Curated data is a representation of reality based on data analysis. It is divided into three categories (training, validation and testing). Validation data is used for hyperparameter tuning, training data is used for parameter optimization for a selected set of hyperparameters, and together they are used to create the trained model. Testing data is used for ``validation'' of the trained model (interestingly validation data is not used for this purpose).

The successful development and deployment of AI/ML models for nuclear reactor applications does not depend only on the application of rigorous VVUQ process but also in addressing some quality assurance aspects of ML models in order to build confidence and trust in the predictions of the models. This is referred to as \textit{ML trustworthiness} and it is a broad area of research that encompasses a few closely related principal subdomains \cite{ali2023explainable}, including \textit{accuracy, robustness (reproducibility, applicability), algorithmic fairness, algorithmic transparency (explainability, interpretability), and privacy/confidentiality}. Among these, four important processes should be naturally incorporated in the VVUQ of ML models, as illustrated in Figure~\ref{figure:VVUQ} and briefly explained below. 

\textit{ML explainability} is the ability to ensure that algorithmic decisions as well as any data driving those decisions can be explained to end-users and stakeholders in non-technical terms. It helps stakeholders and decision makers to understand ML solutions by ``opening the black-box''. ML explainability is important due to its close connection to ML accountability, trustworthiness, compliance, performance and enhanced control. Simpler algorithms such as linear models and decision trees are more explainable than complicated ones, such as neural networks. Explainable AI (XAI) \cite{ali2023explainable, arrieta2020explainable} is an area under active development to enable end users to better understand, trust, and effectively manage ML systems.

\textit{ML interpretability} is the degree to which an ML model obeys structural knowledge of the domain, such as monotonicity, causality, structural constraints, additivity, or physical constraints derived from domain knowledge. Interpretability has usually been used interchangeably with explainability in some literature, but it is believed to be a distinct and domain-specific notion by many others \cite{rudin2019stop}. There has been a debate over whether to (i) build interpretable ML models in the first place, or (ii) to explain black-box ML models. The latter is considered problematic because a second model is usually created to explain the first black box model.

%Simple models such as linear regression and decision trees are more interpretable than complex models such as DNN. This motivates the usage of as simple as possible models, while the usage of more complex ones should be clearly justified. 

\textit{ML reproducibility} is the ability to replicate the ML model from data processing to model design, reporting, model analysis, or evaluation to successful deployment. The usage of ML models for high consequence systems, as it is the case for nuclear reactors, and especially with regards to their licensing, renders reproducibility very important for the ML models deployment. Good reproducibility means the model is scalable and ready to push for production with large-scale deployment, which is critical for ML-based advanced reactor analysis. Standards based on data, model, programming best practices and workflow automation developed in other areas for ML reproducibility such as life science \cite{heil2021reproducibility} and health care \cite{beam2020challenges} can be leveraged into quantifiable criteria in this VVUQ framework.

\textit{ML applicability} is the usability of ML for new scenarios such as unseen domains. For advanced reactors, this is essential because there are issues surrounding the acceptance of ML by stakeholders and decision makers.
Example of these issues are (a) how to evaluate applicability, (b) how to evaluate cost (including re-tooling), (c) how to evaluate improvement and value relative to present human processes, and (d) how to foster user adoption. Quantitative metrics need to be developed, for example, can be based on the domain generalization methods \cite{wang2022generalizing}, including data manipulation (data augmentation/generation), representation learning (domain-invariant representation learning and feature disentanglement), and learning strategy (ensemble learning and meta-learning).

%The validation domain of the ML model should be clearly defined. The extrapolation of the ML model outside of this domain should be identified when performing predictions followed by an assessment of its performance. This shares similarities with the scaling analysis performed for the VVUQ of physical models \cite{dzodzo2019scaling,huang2022pcm}.

%%%%%%%%%%%%%%%%%%%%%%%%%%%%%%%%%%%%%%%%%%%%%%%%%%%%%%%%%%%%%%%%%%%%%%%%%%%%%%%%
\section{Conclusions}
\label{section:conclusions}
%%%%%%%%%%%%%%%%%%%%%%%%%%%%%%%%%%%%%%%%%%%%%%%%%%%%%%%%%%%%%%%%%%%%%%%%%%%%%%%%

The existing applications of SciML in various nuclear engineering problems have often outpaced our formal understanding of the AI/ML algorithms. One important but underrated area is UQ of SciML models. SciML models are subject to approximation uncertainty when they are used to make predictions, due to sources including but not limited to, training data noise, data coverage, extrapolation outside of the training domain, imperfect ML model architecture and stochastic training process. In this paper, we have elucidated the significance of UQ of SciML. We explained the differences in the basic concepts of UQ of physics-based models and data-driven ML models. We discussed, demonstrated, and compared the various sources of uncertainties in physical modeling and SciML. we presented two demonstration examples (analytical GP and SAFARI-1 axial neutron flux profiles) to show the performance of five available approaches for UQ of ML, MCD, DE, BNN, CP and GP. We analyzed the characteristics of these methods based on a few desired properties for the predicted uncertainties. Finally, we briefly discussed the needs to perform VVUQ of ML models. 
UQ is a critical step to establish confidence in the ML model predictions. It should be considered in the pipeline of ML model development process to help build/improve the ML credibility. The nuclear community should put more efforts into quantification of ML uncertainties in order to facilitate more trustworthy ML applications in high-consequence nuclear systems. it is important to note that most of these discussed quality assurance aspects are currently under consideration by the U.S. NRC \cite{dennis2023artificial} and U.S. Department of Energy laboratories \cite{muhlheim2023status}.

%%%%%%%%%%%%%%%%%%%%%%%%%%%%%%%%%%%%%%%%%%%%%%%%%%%%%%%%%%%%%%%%%%%%%%%%%%%%%%%%
\newpage
\section*{CRediT authorship contribution statement}
%%%%%%%%%%%%%%%%%%%%%%%%%%%%%%%%%%%%%%%%%%%%%%%%%%%%%%%%%%%%%%%%%%%%%%%%%%%%%%%%

\textbf{X.~Wu}: Conceptualization, Methodology, Writing - original draft,
\textbf{L.~E.~Moloko}: Methodology, Investigation, Visualization, Writing -- review \& editing
\textbf{P.~M.~Bokov}: Methodology, Investigation, Visualization, Writing -- review \& editing
\textbf{G.~K.~Delipei}: Methodology, Investigation, Visualization, Writing -- review \& editing, 
\textbf{J.~Kaizer}: Writing -- review \& editing, 
\textbf{K.~N.~Ivanov}: Writing -- review \& editing.

%%%%%%%%%%%%%%%%%%%%%%%%%%%%%%%%%%%%%%%%%%%%%%%%%%%%%%%%%%%%%%%%%%%%%%%%%%%%%%%%
\section*{Declaration of Competing Interest}
%%%%%%%%%%%%%%%%%%%%%%%%%%%%%%%%%%%%%%%%%%%%%%%%%%%%%%%%%%%%%%%%%%%%%%%%%%%%%%%%

The authors declare that they have no known competing financial interests or personal relationships that could have appeared to influence the work reported in this paper.

%%%%%%%%%%%%%%%%%%%%%%%%%%%%%%%%%%%%%%%%%%%%%%%%%%%%%%%%%%%%%%%%%%%%%%%%%%%%%%%%
%\newpage
\section*{Acknowledgments}
%%%%%%%%%%%%%%%%%%%%%%%%%%%%%%%%%%%%%%%%%%%%%%%%%%%%%%%%%%%%%%%%%%%%%%%%%%%%%%%%

This work was funded by the U.S. Department of Energy (DOE) Office of Nuclear Energy Distinguished Early Career Program (DECP) under award number DE-NE0009467. Any opinions, findings, and conclusions or recommendations expressed in this paper are those of the authors and do not necessarily reflect the views of the U.S. DOE.

%%%%%%%%%%%%%%%%%%%%%%%%%%%%%%%%%%%%%%%%%%%%%%%%%%%%%%%%%%%%%%%%%%%%%%%%%%%%%%%%
%%%%%%%%%%%%%%%%%%%%%%%%%%%%%%%%%%%%%%%%%%%%%%%%%%%%%%%%%%%%%%%%%%%%%%%%%%%%%%%%
%%\section*{References}
\newpage
\bibliography{./bibliography.bib}

\end{document}